\documentclass[a4paper,11pt]{article}
\usepackage{subcaption}

\pdfoutput=1 

\usepackage{jinstpub} 
                     
\usepackage{lineno}
\usepackage{multirow}
\usepackage{graphicx}

\title{A comparison of the neutron detection efficiency and response characteristics of two pixelated PSD-capable organic scintillator detectors with different photo-detection readout methods}

\author[a,1]{J. Balajthy\note{Corresponding author.}}
\author[a]{P. Marleau}
\author[a]{M. Sweany}

\affiliation[a]{Sandia National Laboratory, Livermore, CA 94550, USA}
\emailAdd{jabalaj@sandia.gov}

\abstract{We characterize the performance of two pixelated neutron detectors: a PMT-based array that utilizes Anger logic for pixel identification and a SiPM-based array that employs individual pixel readout. 
The SiPM-based array offers improved performance over the previously developed PMT-based detector both in terms of uniformity and neutron detection efficiency. 
Each detector array uses PSD-capable plastic scintillator as a detection medium.
We describe the calibration and neutron efficiency measurement of both detectors using a $^{137}$Cs source for energy calibration 
and a $^{252}$Cf source for calibration of the neutron response.
We find that the intrinsic neutron detection efficiency of the SiPM-based array is ($30.2 \ \pm \ 1.7$)\%, which is almost twice that of the PMT-based array, which we measure to be ($16.9 \pm 0.2$)\%.
}


\begin{document}
\maketitle
\flushbottom

\section{Introduction}

Enabling a fieldable pixelated fast neutron detection or imaging plane has many challenges, including maximizing the neutron detection efficiency 
while minimizing size, weight, and power (SWaP).
Some current and widely used fast neutron detection systems minimize 
the number of readout channels, and thus system complexity and power requirements, by employing Anger logic readout using photomultiplier tubes
\cite{anger}. However, this leads to poor light collection efficiency for certain pixels and does not allow for multiple interactions from the 
same neutron to be resolved, which reduces the intrinsic neutron detection efficiency \cite{newby2}.  
Advances in Silicon Photomultiplier (SiPM) technology makes highly pixelated photo-detection with $O(100)$ ps 
timing feasible, allowing for cost effective independent readout of scintillator pixels and the ability to resolve multiple fast neutron scatters in the same volume that occur 
within less than 1~ns. Furthermore, the improved photo-detection efficiency of SiPMs compared to PMTs \cite{zappala,nakamura,romeo}
coupled with light collection
efficiency improvements from 1-to-1 scintillator/photodetector coupling has the potential to greatly improve the intrinsic neutron detection 
efficiency. SiPMs are also lighter, more compact, and require less power than PMTs. Provided enough investment into low 
power, high channel acquisition using either FPGA-based methods (see e.g. \cite{zhao1, zhao2, zheng}) or ASICs, a pixelated fast neutron 
detection plane with SiPM-based readout of each scintillator pixel may approach the necessary efficiency and SWaP requirements for fieldable 
systems. 

Here, we quantify the intrinsic neutron detection efficiency of two pixelated neutron detectors: one 10x10 pixelated array read out with four 
PMTs via Anger logic, and one 8x8 pixelated array individually read out with a commercial 8x8 SiPM array. The two detectors differ slightly in
scintillator composition, however the largest differences are the size and number of scintillator pixels and how they are read out.
We use the nomenclature "PMT-based" and "SiPM-based" to distinguish
between the two. Both scintillator arrays are manufactured by Agile Technologies. Section~\ref{sec:exp} describes the two detectors, and
also details the measurement setup used to characterize them.  Section~\ref{sec:res} presents the calibration and response of each detector. Finally, the 
intrinsic neutron detection efficiency results are presented in Section \ref{sec:eff}. 

\section{Experimental Setup}
\label{sec:exp}

\begin{figure}[!htbp]
	\centering
	\includegraphics[angle=0,width=0.6\columnwidth]{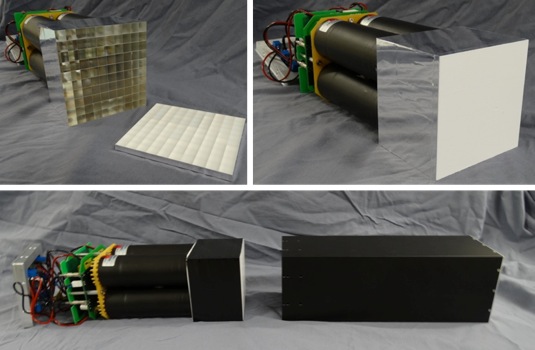}
	\caption{Images of the PMT-based detector, courtesy P. A. Hausladen of Oak Ridge National Laboratory.  }
	\label{fig:fig_olddetectors} 
\end{figure}

The PMT-based detector consists of one hundred 1.08 cm x1.08 cm x 5 cm
pixels of pulse shape discrimination (PSD)-capable plastic scintillator EJ299-34.
Each scintillator pixel is individually wrapped with a layer of 3M's 
Enhanced Specular Reflector  (ESR) material.  The 10x10 pixelated scintillator array is read out 
by four Hamamatsu R7724-100 super bialkali photomultiplier tubes, 
and Anger logic is used to determine which pixel a scintillation event 
occurred in. An acrylic (PMMA) light guide couples the scintillator array to the four PMTs.
The detectors also include a layer of $^6$Li/ZnS:Ag for thermal neutron detection.  
ZnS:Ag has a phosphorescence response to the particles ($\alpha, t$) released 
by neutron capture on $^6$Li.  Because of its comparatively long decay 
time, phosphorescence is easily distinguished from scintillation in EJ-299-34 
by PSD. A photograph of a single detector is shown in Figure \ref{fig:fig_olddetectors}. 
The performance of the PMT-based detectors has been 
characterized in previous work \cite{newby1, newby2}, however 
in order to provide a comparison with the SiPM-based detector using the same characterization
procedures, much of the characterization is repeated here. 

\begin{figure}[!htbp]
	\centering
	\begin{subfigure}[h]{0.22\columnwidth} 
		\includegraphics[angle=0,width=\columnwidth]{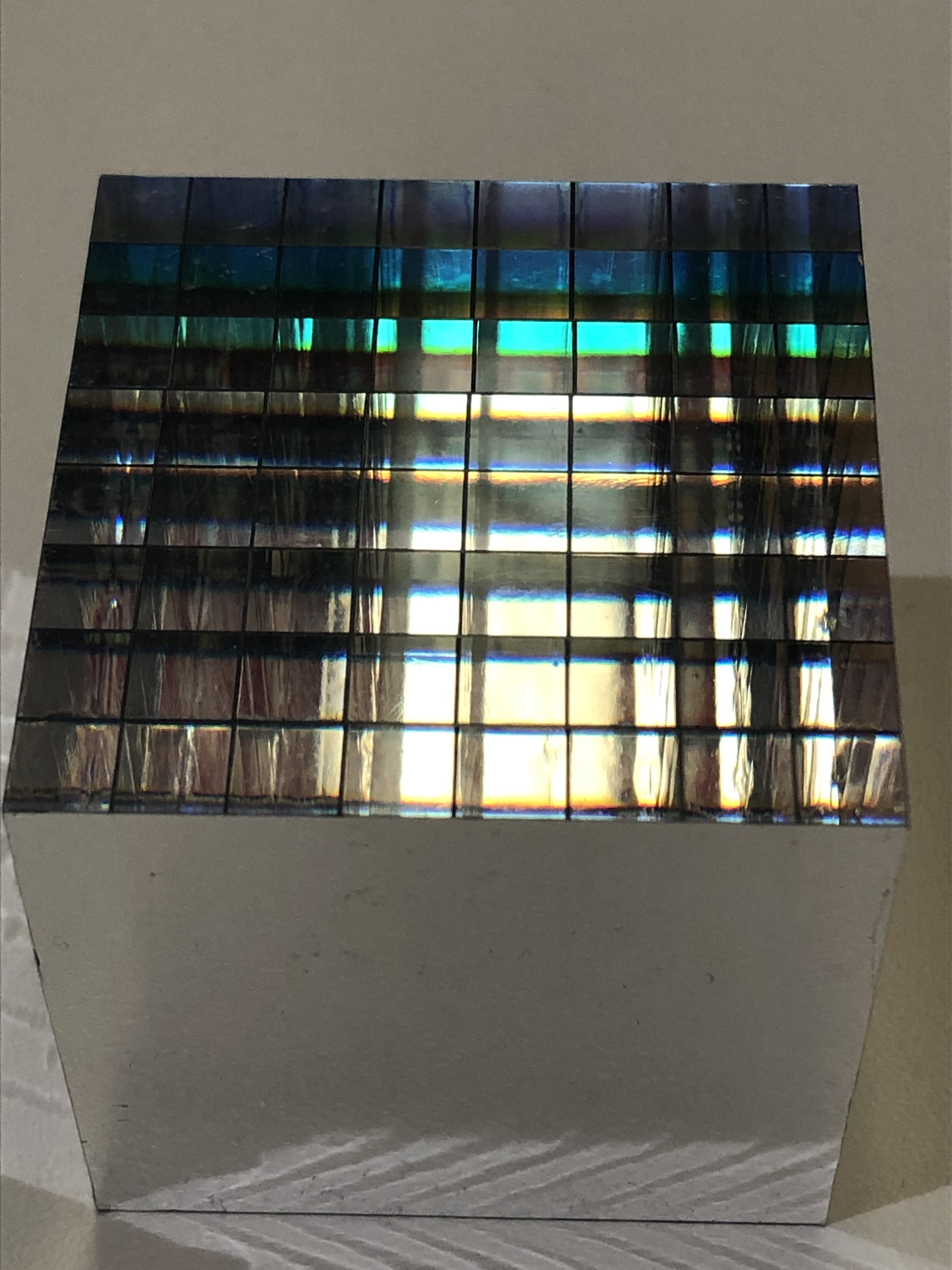}
		\caption{}
	\end{subfigure}	
	\begin{subfigure}[h]{0.22\columnwidth} 
		\includegraphics[angle=0,width=\columnwidth]{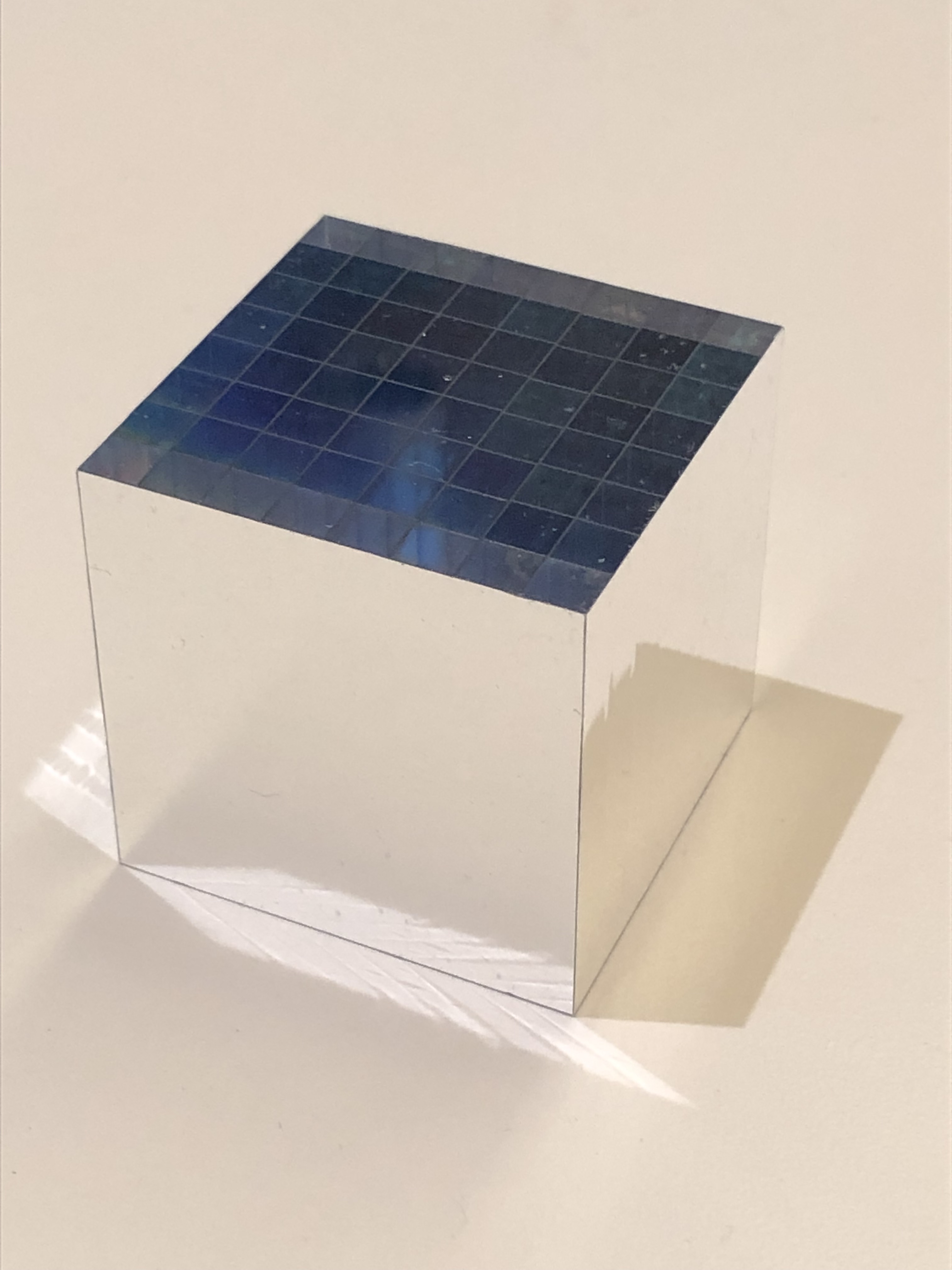}
		\caption{}
	\end{subfigure}
	\begin{subfigure}[h]{0.39\columnwidth} 
			\includegraphics[angle=0,width=\columnwidth]{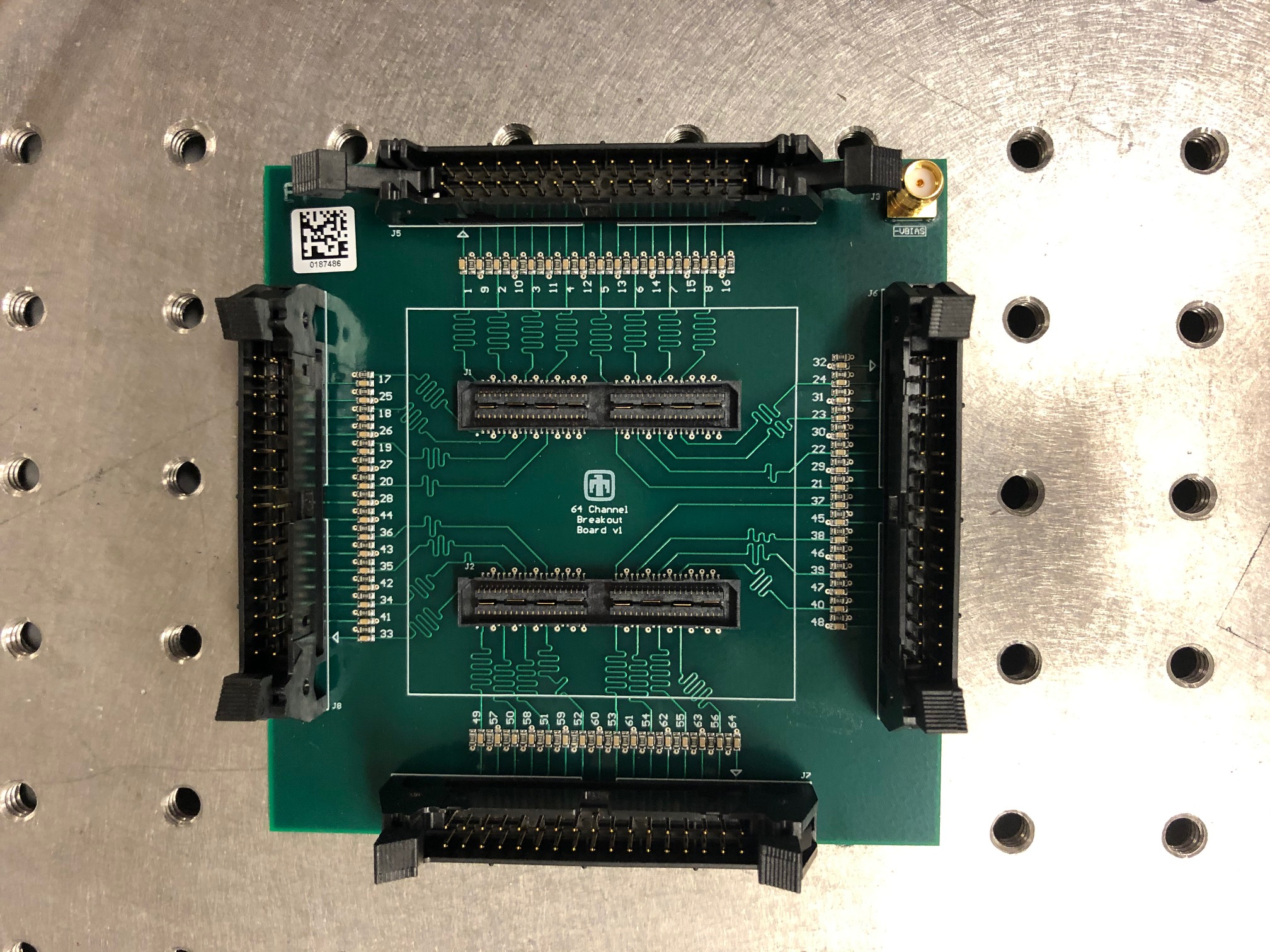}
			\caption{}
		\end{subfigure}	

	\caption{The SiPM-based scintillator array (a and b), and the custom SiPM readout board (c).}
	\label{fig:fig_newdetectors} 
\end{figure}

The SiPM-based detector is an 8x8 array of EJ-299-33M PSD-capable 
plastic scintillator, and is shown in Figure \ref{fig:fig_newdetectors}a,~b. The scintillator pixel size and pitch are designed to match the SiPM array
ArrayJ-60035-64P-PCB from SensL/onsemi, which has 6.13 mm x 6.13 mm SiPMs with a pitch of 6.33 mm mounted
on a printed circuit board (PCB). The scintillator pixels are 5~cm long. Between 
each scintillator pixel are two layers of 3M's ESR and one layer of Aluminum between the two ESR layers. 
ESR is also layered on the front side of the array, which does not include $^6$Li/ZnS:Ag. The scintillator array is coupled with
silicon-based optical grease directly to the SiPM array without a light guide. The SiPM array 
is mated to a custom 64 channel breakout board, shown in Figure \ref{fig:fig_newdetectors}c, that 
is based on the readout circuit described in \cite{giha}. The board is biased at (-)30 V 
with a B\&K Precision 1761 DC power supply. The standard outputs of the SiPM pixels are coupled in groups 
of 16 with header pins that mate to CAEN's A385 adaptor with MCX connectors.  
The assembly is housed in a light tight enclosure. 
		
A $^{137}$Cs source is used for energy calibration for both the PMT-based detector and the SiPM-based detector. The $^{137}$Cs
source is positioned such that it is approximately centered with respect to the face of the detector array, at a distance such that the
rate is maximized without overwhelming the data acquisition (DAQ). This is approximately 40~cm for the PMT-based detector, and 10~cm for the
SiPM-based array. 

A single $^{252}$Cf measurement for each of the detectors is used for both the PSD calibration and neutron efficiency
measurement. For each of the measurements, the $^{252}$Cf source is positioned so that it is centered to within 1~cm with respect
to the face of the detector. For the PMT-based detector measurements, the source is placed at a distance of 142.0~$\pm$~0.3~cm from the face
of the detector, and for the SiPM-based measurement, the source is placed at 40.0~$\pm$~0.3~cm from the face of the detector. 

Each PMT anode output is passed through an integrating preamplifier \cite{newby1} before being digitized using a CAEN DT5730s desktop digitizer
with DPP\_PSD firmware \cite{dt5730}. The digitizer samples at rate of 500~MHz and is operated with a 2V vertical scale.
The waveforms from the PMTs are acquired using the CoMPASS data acquisition software from CAEN \cite{compass}. 
The local triggers for each PMT channel are configured in the leading edge mode.
The digitizer is configured so that a local trigger from any of the four PMT channels will initiate a global trigger, 
and all four PMT channels are read out simultaneously on any global trigger. 
There are a small number of events (<0.04\%) in which the waveforms for individual channels have been lost, likely due to pileup. 
This number of events will be negligible for our efficiency calculations in Section~\ref{sec:eff}.

Waveforms from the 64-channel SiPM array are digitized using four CAEN v1725 digitizers with DPP\_PSD firmware, and installed in a VME crate \cite{v1725}. 
The v1725 digitizers have a sample rate of 250~MHz and are operated with a 500 mV full-scale and a 5\% DC offset. 
The CoMPASS software from CAEN is again used for waveform acquisition. We configure the digitizers to acquire asynchronously, allowing each channel to 
trigger independently. The channel triggers are all configured in ``CFD'' mode, with a 28~ns CFD delay and a 50\% CFD fraction. 
The CFD threshold is set to~110~lsb for most channels, or roughly 0.050~MeVee for neutrons. 
Several channels are observed to have excess electronic noise, and their thresholds were 
set higher. No triggers were reported as being lost by the CoMPASS software in the SiPM-based $^{137}$Cs and $^{252}$Cf measurements. 

\section{Data Processing and Calibration}
\label{sec:res}

Data processing and calibration operations are mostly written in C++, utilizing algorithms from the ROOT analysis toolkit \cite{root}.  Some high-level 
analysis is performed in Python, including the time offset calibration shown in Figure~\ref{fig:sipmcoinpeaks} and the neutron efficiency calculations 
described in Section~\ref{sec:eff}. The Python packages we use most heavily are NumPy\cite{numpy}, SciPy\cite{scipy}, MatPlotLib\cite{mpl}, and upRoot\cite{uproot}. 
Digital waveform processing is applied to pulses from both the PMT-based and SiPM-based detectors and are then 
organized by time and pixel number. The processed pulse values are then calibrated for energy and PSD. The calibrated data is used in 
Section~\ref{sec:eff} for the neutron efficiency calculation.

\begin{figure}[!htbp]
	\centering
	\includegraphics[angle=0,width=0.5\columnwidth]{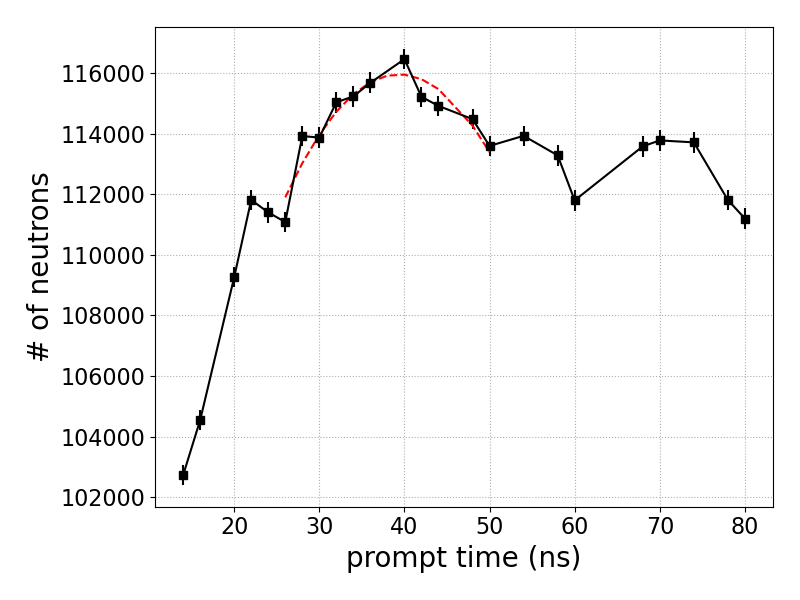}
	\caption{ Optimization for prompt gate time in the PMT-based detector. The black markers show the number of observed neutrons as a function of prompt gate time. The red dashed line shows a quadratic fit to these results. Results are calculated using about 10\% of the entries in the $^{252}$Cf dataset.}
	\label{fig:blockpromptopt} 
\end{figure}

\subsection{PMT-based Detector Processing}

Waveforms from the four channels of the PMT-based detector are baseline-subtracted using samples 10 through 60 and are then sorted and matched according 
to the fine timestamp reported by the digitizer. 
For a given interaction, the four matched waveforms are summed together in software, and the summed 
waveform is smoothed using a 5-bin centered moving average filter. The pulse time relative to the trigger window is taken to be 
the linearly interpolated value between samples in the summed and smoothed waveform corresponding to 50\% of the maximum value on the rising edge. 
The individual PMT waveforms are smoothed using the same 5-bin centered moving average filter, are baseline-subtracted, and then are linearly interpolated 
at $t = 40$~ns and $t = 832$~ns relative to the pulse time.
These values are stored for all four PMTs as the prompt and total pulse amplitudes, respectively ($A_p$ and $A_t$).
A gate time of 832~ns for $A_t$ has been previously identified as yielding optimal energy results \cite{newby1, newby2}. 
We optimize the gate time for $A_p$ using approximately the first 10\% of entries in the $^{252}$Cf. We process this subset of data 
multiple times, using a different prompt gate time for each iteration. 
We calculate the number of observed neutrons for each processing as described in Section~\ref{sec:eff}, and fit a quadratic polynomial to the values around the maximum, as is shown in Figure~\ref{fig:blockpromptopt}. The optimal prompt gate time is taken to be the maximum value of the quadratic fit.

\begin{figure}[!htbp]
	\centering
	\includegraphics[angle=0,width=0.5\columnwidth]{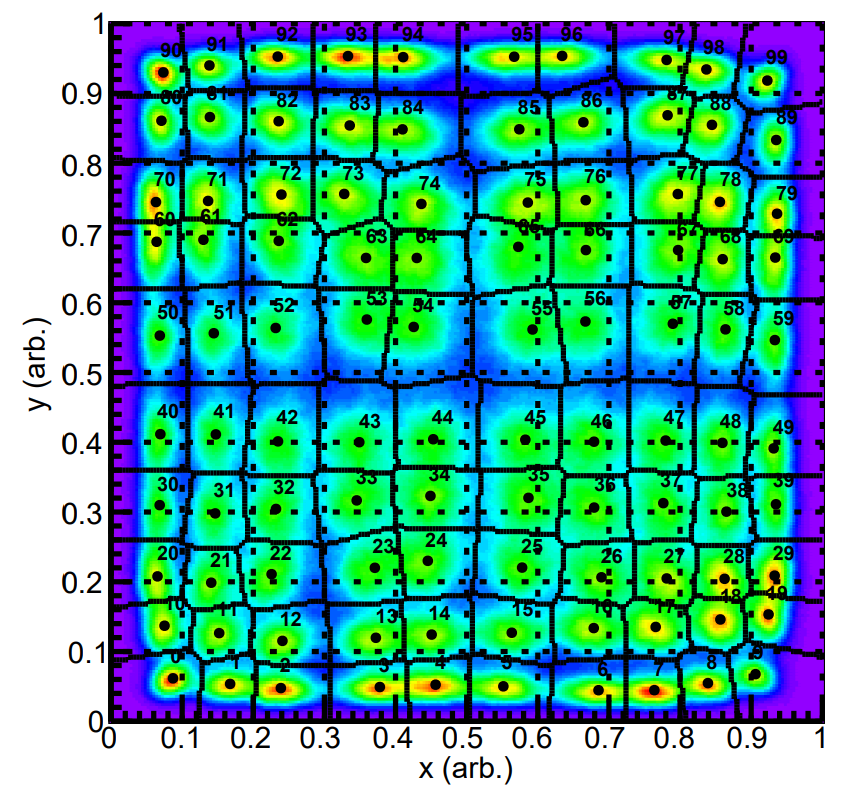}
	\caption{Anger logic look-up table for the PMT-based detector, measured using the $^{252}$Cf . The values of x  and y are values of the 
Anger logic $x$ and $y$ described in the text. The solid black lines show the divisions between the detector pixels, and 
the dashed lines show a uniform 10x10 grid. The numbered markers indicate the peak position for each of the 100 detector pixels.} 
	\label{fig:angerlut} 
\end{figure}

The $^{252}$Cf measurement for the PMT-based detector is used to generate a look up table to identify the pixel number from the $x$-position and 
$y$-position measures from Anger logic:

\begin{linenomath*}
\begin{equation}
x_i = \frac{A_{2,i}+A_{3,i}}{\displaystyle\sum_{p=0}^3A_{p,i}} \text{,   and   }  y_i = \frac{A_{0,i}+A_{2,i}}{\displaystyle\sum_{p=0}^3A_{p,i}}
\end{equation}
\end{linenomath*}
For the $i^{th}$ entry in a measurement, $A_{p,i}$ is the ``total'' pulse integral value observed by the PMT with index $p$. The uppermost PMTs have 
indices $p=0$ and $p=2$, while the leftward most PMTs have indices $p=2$ and $p=3$. The software for the implemenation of the look up table algorithm was developed 
by M. A. Blackston and R. J. Newby of Oak Ridge National Laboratory. The algorithm seeds 100 2-D Gaussian fits based on local maxima with an assumed standard deviation for each pixel.
The cell boundaries are determined based on where neighboring pixels' responses are equalized. A human user checks and refines the result if necessary. 
The measured Anger logic lookup table is shown in Figure~\ref{fig:angerlut}. 

\subsection{SiPM-Based Detector Processing}

The waveforms from the SiPM-based detector are baseline subtracted using the first 80 ns of the trace and a centered moving average filter 
is applied.  The cumulative distribution of the baseline-subtracted waveform is used to calculate the pulse integral for total energy deposition 
measurements. The integral is evaluated exactly 280~ns after the measured pulse time using a linear interpolation of 
the cumulative distribution of the waveform. The pulse time is the linearly interpolated value 
between waveform samples corresponding to 50\% of the maximum value on the rising edge.
For pulse shape measurements on the SiPM-based detector, we linearly interpolate between samples of the cumulative distribution of the waveform for the 
value corresponding to the desired integration window. 

\begin{figure}[!htbp]
	\centering
	\begin{subfigure}[h]{0.3\columnwidth} 
		\includegraphics[angle=0,width=\columnwidth]{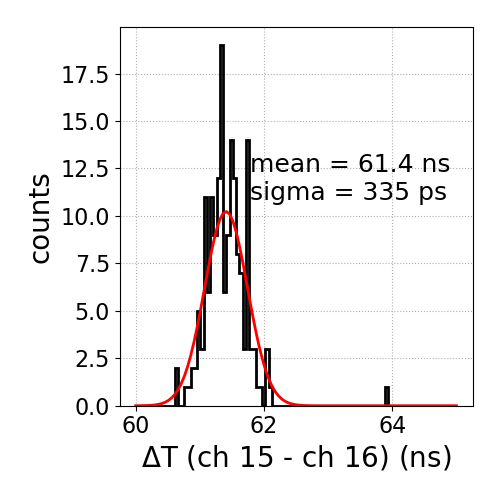}
		\caption{}
	\end{subfigure}	
	\begin{subfigure}[h]{0.3\columnwidth} 
		\includegraphics[angle=0,width=\columnwidth]{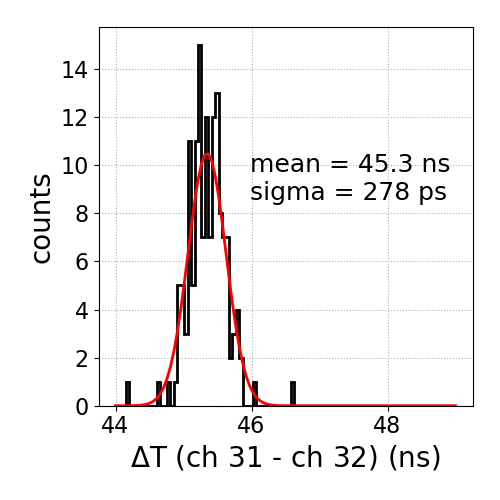}
		\caption{}
	\end{subfigure}
	\begin{subfigure}[h]{0.3\columnwidth} 
			\includegraphics[angle=0,width=\columnwidth]{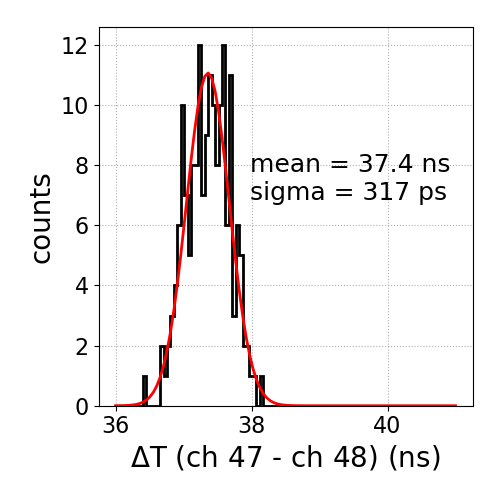}
			\caption{}
		\end{subfigure}	

	\caption{Coincidence peaks for $^{22}$Na calibration of SiPM array timing offsets. Plot (a) shows the time differences for digitizer 
boards 0 and 1, (b) shows the time differences for boards 1 and 2, and (c) shows the time differences for boards 2 and 3.}
	\label{fig:sipmcoinpeaks} 
\end{figure}

The fine timestamp reported by the CAEN digitizers is used for the trigger time of each interaction. Time offsets between the four v1725 
boards are measured using $^{22}$Na coincident interactions. Two 6 mm x 6mm x 6mm {\it trans}-Stilbene crystals are coupled to SiPM 
pixels that are read out by neighboring 
digitizer boards, and the $^{22}$Na source is placed directly between the two. The time difference spectrum is measured for each of the 
three selected channel pairs (15 and 16, 31 and 32, and 47 and 48), and the coincidence peak is identified. We use only events with pulse 
heights between 150~and~350~mV for our coincidence analysis. The location of the $^{22}$Na Compton edge ranges 
from 254~to~290~mV for the six pixels tested, so the pulse height selection cut corresponds to roughly 190~keVee to 450~keVee. 
The Gaussian mean and width of each coincidence peak is shown in Figure~\ref{fig:sipmcoinpeaks}. We measure a coincident 
timing resolution of about 300~ps for each channel pair. 
The interaction timestamps for the $^{137}$Cs and $^{252}$Cf datasets are adjusted according to the mean values of the $^{22}$Na 
coincident peaks.
We will refer to the adjusted fine time stamp for interaction $i$ in a given dataset as $T_i$.

A coincidence rejection processing step is added for the SiPM-based $^{137}$Cs and $^{252}$Cf datasets. This primarily rejects 
pulses caused by light sharing between detector pixels and pulses due to multiple scattering interactions in adjacent pixels 
by the same particle. This also removes accidental coincidences 
from our dataset, which is expected to be small compared to light sharing and multi scatter events. 

We identify ``events'' of coincident pulses by looping across all trigger times in a dataset. A pair of consecutive pulses, $i$ and $i+1$, 
is considered coincident and both are elements of the same event, $N$, if $T_{i+1}-T_{i} < 50$ ns. In the case that $T_{i+2}-T_{i+1}$ is also 
$< 50$ ns, then pulse $i+2$ will also be included in event $N$. Subsequent pulses are added to event $N$ until a pair of non-coincident 
pulses is found. If a pulse $i$ is coincident with neither $i+1$ or $i-1$, it will be assigned to its own event with only one pulse. 
Events will therefore range in size from a single pulse to an 
arbitrary number of coincident pulses. For each event we label the pulse with largest integral value as the ``primary'' interaction, 
and we label the rest of the pulses in the event as ``coincidence'' interactions. We expect this to be sufficient for neutron efficiency measurements, as the order of pulses in an event is not important, 
however a more careful treatment is required for neutron imaging to identify the pixel location of the first pulse in a chain.
The set of pulses identified as primary interactions are used for the energy calibration, PSD calibration, and neutron detection efficiency calculations.

\begin{figure}[!htbp]
	\centering
	\begin{subfigure}[h]{0.4\columnwidth} 
		\includegraphics[angle=0,width=\columnwidth]{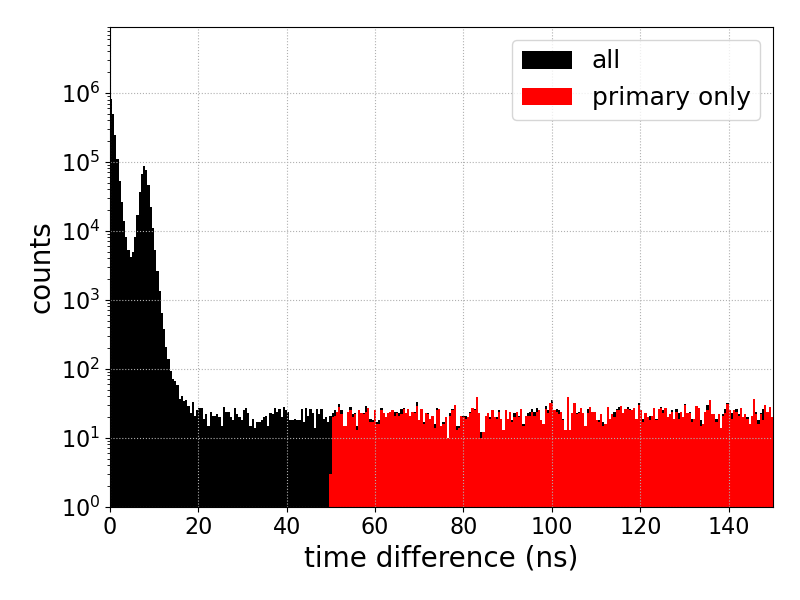}
		\caption{}
	\end{subfigure}	
	\begin{subfigure}[h]{0.4\columnwidth} 
		\includegraphics[angle=0,width=\columnwidth]{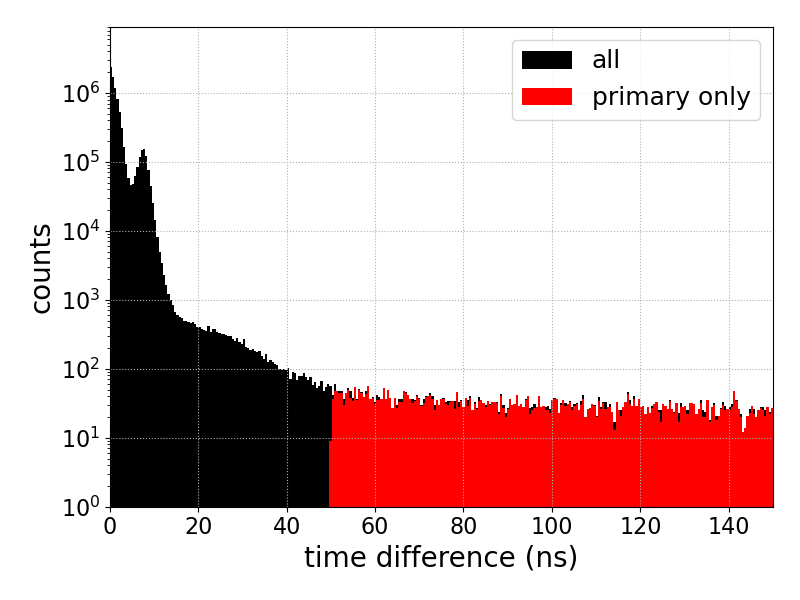}
		\caption{}
	\end{subfigure}

	\caption{Histograms of time differences between consecutive interactions for the SiPM-based $^{137}$Cs (left) and $^{252}$Cf datasets (right). 
	The black histograms contain all interactions in the datasets, while the red histograms contain only interactions identified as ``primary''  by the 
	coincidence rejection described in the text.}
	\label{fig:tdiffhist} 
\end{figure}

We select a time difference threshold of 50~ns in order to reject all the excess low-time difference pulses seen in the coincidence histograms shown in 
Figure~\ref{fig:tdiffhist}. There are two peaks below 10 ns time difference observed in both the $^{137}$Cs and $^{252}$Cf coincidence 
histograms which we identify as coming primarily from light sharing. We have found that the second of these peaks is due to a mis-calibration 
in the timing offset for the fourth board, likely caused by excess electronic noise, that causes an 8 ns shift in trigger times relative to other boards. 
This shift is well under our 50 ns threshold, so we do not expect it to significantly impact our coincidence rejection. There is a third, broader 
peak that is observed only in the $^{252}$Cf histogram that ends at about 50 ns time difference that we attribute to multiple scatters of
fast neutrons.

\begin{figure}[!htbp]
	\centering
	\begin{subfigure}[h]{0.4\columnwidth} 
		\includegraphics[angle=0,width=\columnwidth]{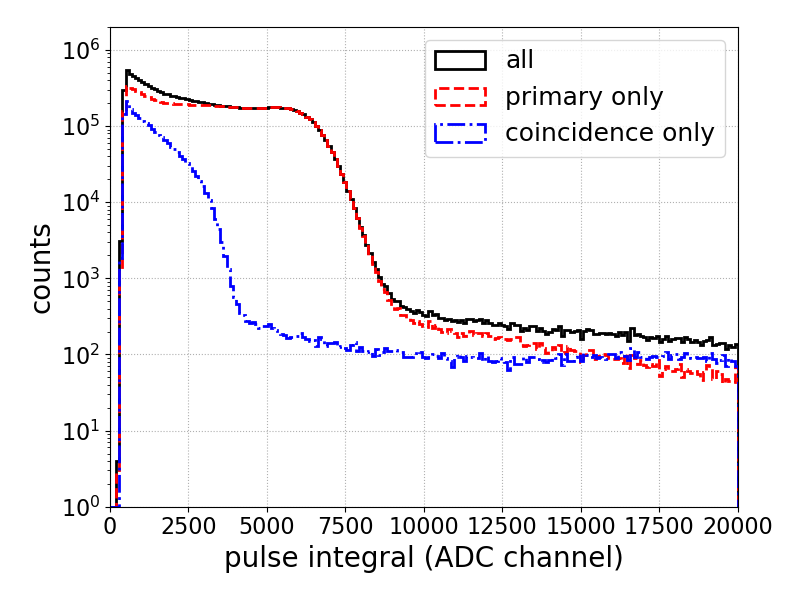}
		\caption{}
	\end{subfigure}	
	\begin{subfigure}[h]{0.4\columnwidth} 
		\includegraphics[angle=0,width=\columnwidth]{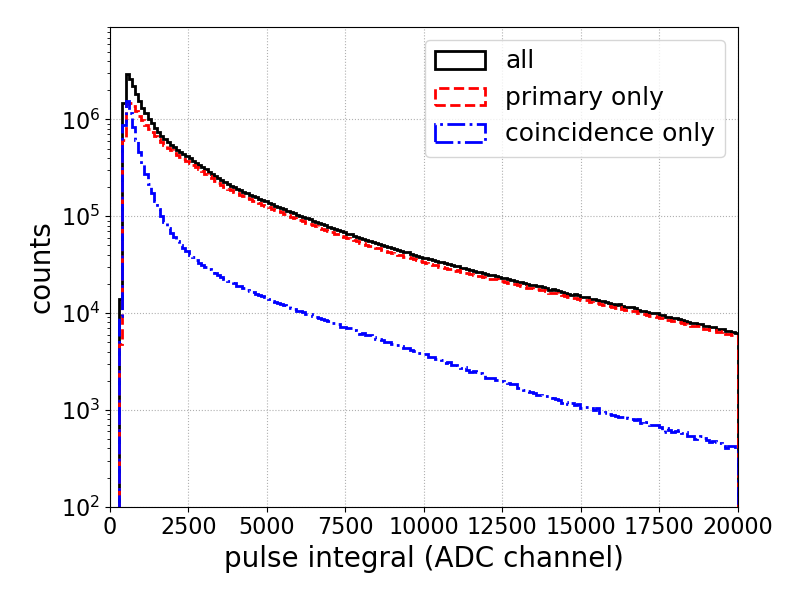}
		\caption{}
	\end{subfigure}

	\caption{Histograms of pulse integrals for the SiPM-based $^{137}$Cs (left) and $^{252}$Cf datasets (right). The black histograms contain all events in the datasets, the red histograms contain only events identified as ``primary'' by the coincidence rejection described in the text, and the blue histograms contain only events labeled ``coincidence''.}
	\label{fig:ihistscoinvprim} 
\end{figure}

The effects of the coincidence rejection on the pulse integral spectra for the the $^{137}$Cs and $^{252}$Cf datasets are shown in Figure~\ref{fig:ihistscoinvprim}. 
The ``primary'' only selections show a significant reduction at low pulse integrals. The Compton edge of the $^{137}$Cs spectrum appears to be well preserved, and 
the shape of the $^{252}$Cf spectrum above ADC channel 2500 is largely unaffected. The coincidence rejection removes 
14.9\% of events from the $^{137}$Cs dataset and 24.4\% of events from the $^{252}$Cf dataset.

\subsection{Energy}

\begin{figure}[!htbp]
	\centering
	\includegraphics[angle=0,width=0.4\columnwidth]{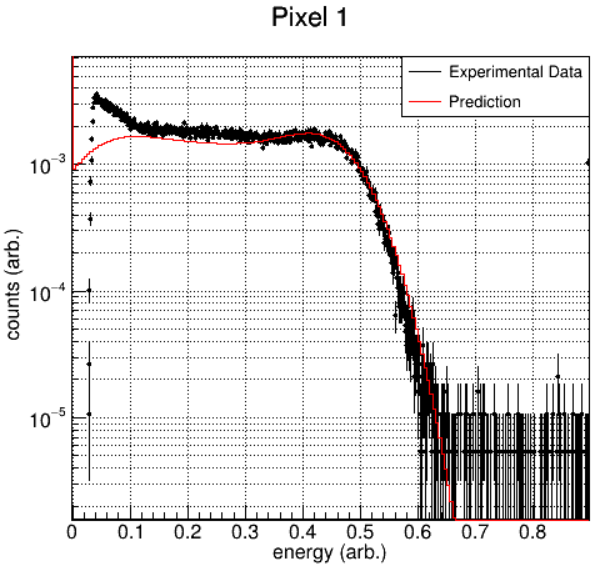}
	\caption{An example Compton edge fit, with the smeared Klein-Nishina spectrum from $^{137}$Cs shown in red, and the 	
	experimental data in black.}
	\label{fig:figexample} 
\end{figure}

The calibration from pulse integral to MeVee is determined by fitting the measured spectrum from a $^{137}$Cs source to a 
Klein-Nishina convolved with an energy-dependent Gaussian \cite{klein}. The method is similar to that described in 
\cite{sweany}. The pulse integral spectrum from the $^{137}$Cs source is used to compare to the expected 
spectrum calculated from the Klein-Nishina prediction. First, the experimental result is converted to MeVee units with a linear transformation:

\begin{linenomath*}
\begin{equation}
E_{MeVee} = q_0E_{mV}.
\end{equation}
\end{linenomath*}
Next, an arbitrary y-scale, $q_1$, is applied to the Klein-Nishina prediction for the gamma emissions, and a Gaussian convolution is performed with a standard deviation of:

\begin{linenomath*}
\begin{equation}
\sigma = q_2E_{MeVee} .
\label{eq:res}
\end{equation}
\end{linenomath*}
The fit region is restricted to a range near the Compton edge. 
In this case, the primary parameter of interest is the conversion factor to MeVee. 
Figure \ref{fig:figexample} shows an example fit for one pixel of the SiPM-based detector.

\begin{figure}[!htbp]
	\centering
	\includegraphics[angle=0,width=0.9\columnwidth]{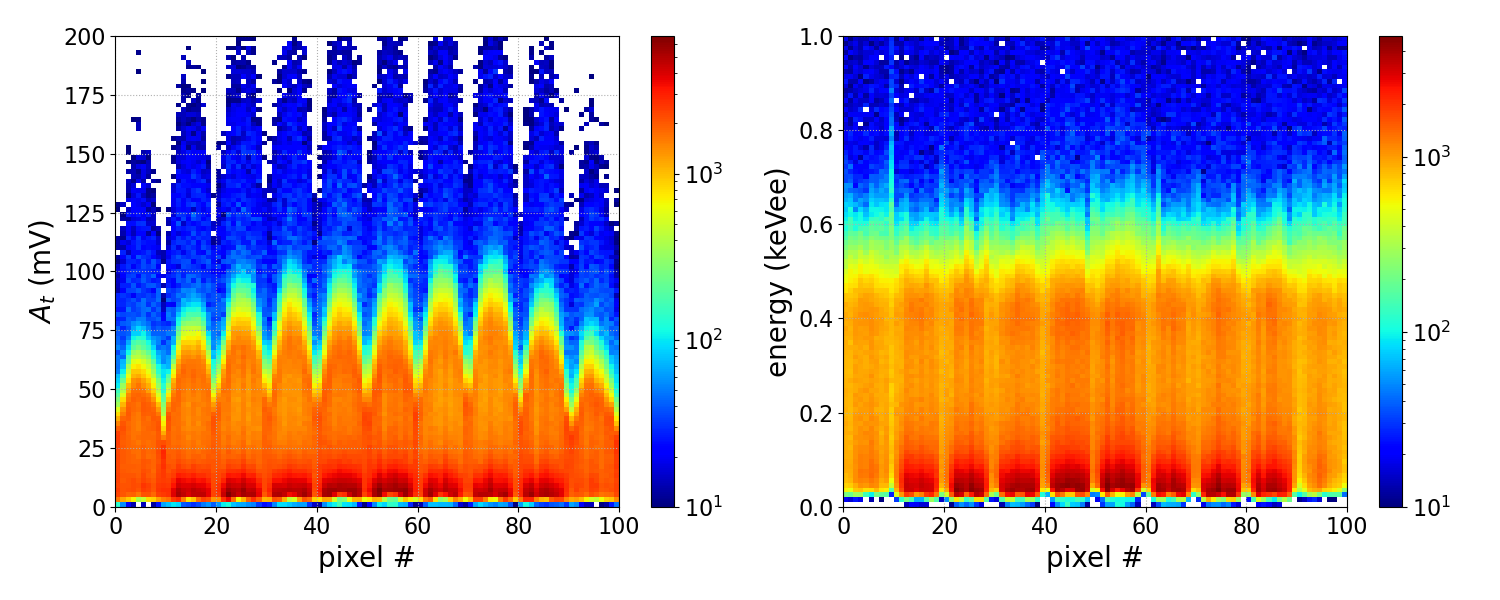}
	\caption{This figure shows the energy response of the PMT-based detector to the $^{137}$Cs calibration. The left panel shows the pulse integral measure ($A_{total}$) for each pixel in the PMT-based detector. The right panel shows the calibrated energy for each pixel in the PMT-based detector. }
	\label{fig:blockenevpix} 
\end{figure}

\begin{figure}[!htbp]
	\centering
	\includegraphics[angle=0,width=0.9\columnwidth]{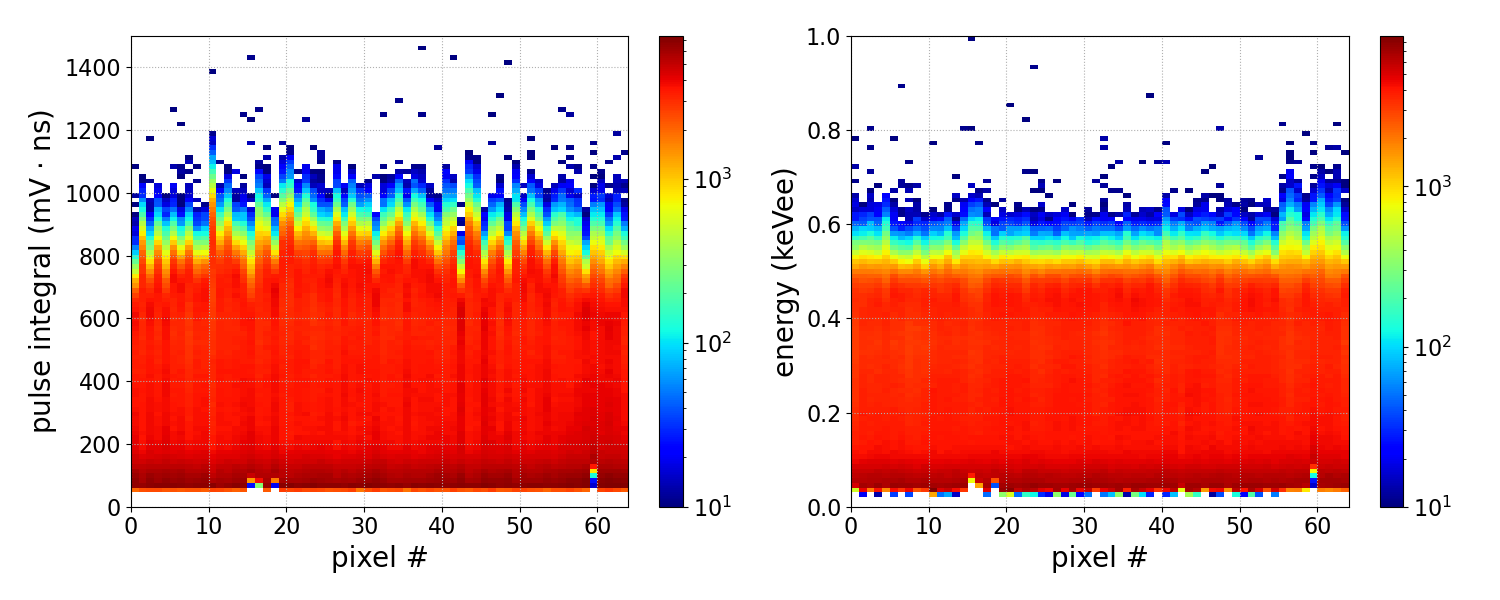}
	\caption{This figure shows the energy response of the SiPM-based array to the $^{137}$Cs calibration. The left panel shows the pulse integral measure ($A_{total}$) for each pixel in the array. The right panel shows the calibrated energy for each pixel in the array. }
	\label{fig:sipmenevpix} 
\end{figure}

\begin{figure}[!htbp]
	\centering
	\begin{subfigure}[h]{0.4\columnwidth} 
		\includegraphics[angle=0,width=\columnwidth]{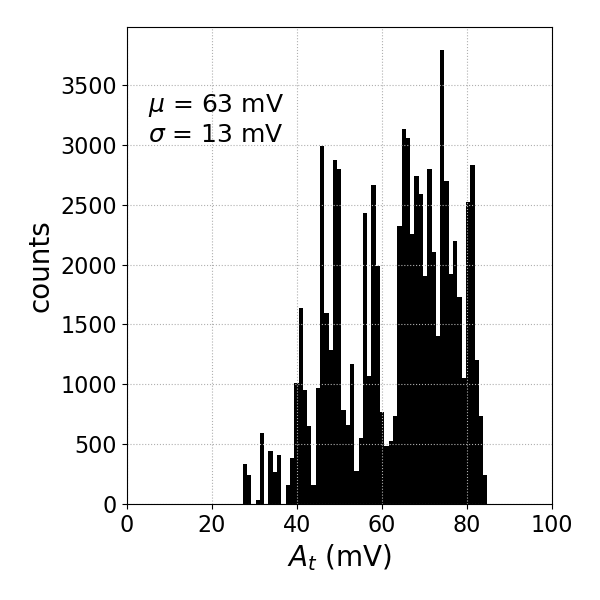}
		\caption{}
	\end{subfigure}	
	\begin{subfigure}[h]{0.4\columnwidth} 
		\includegraphics[angle=0,width=\columnwidth]{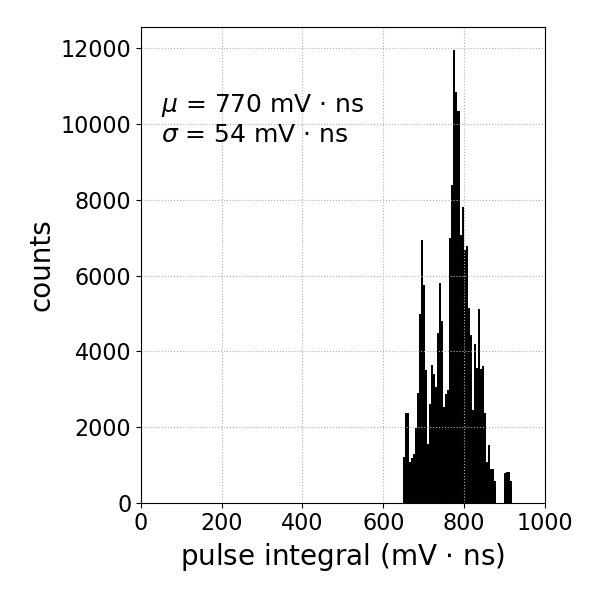}
		\caption{}
	\end{subfigure}

	\caption{This figure compares the light collection uniformity for the PMT-based detector (a) with the SiPM-based array (b). The both plots include pulses with calibrated energy between 0.472-0.482~MeVee from all pixels in the histograms. The PMT-based detector plot uses the total amplitude, converted to mV as the measure of light collection, and the SiPM-based array plot uses pulse integral value in mV$\cdot$ns. }
	\label{fig:lightcolhists} 
\end{figure}

The energy spectra for the calibrated $^{137}$Cs datasets measured by the PMT-based detector and SiPM-based array are shown in 
Figures~\ref{fig:blockenevpix} and~\ref{fig:sipmenevpix}, respectively.  In Figure~\ref{fig:lightcolhists} we compare the integrated 
light collection observed by the two detectors for events with calibrated energy from 0.472-0.482~MeVee.
The total pulse amplitude converted in mV is used as the measure of light collection for the PMT-based detector, and the pulse integral in mV$\cdot$ns
is used for the SiPM-based array. The histograms in Figure~\ref{fig:lightcolhists} contain pulse information from 
all the pixels in the respective detectors. The reported light collection is 63$\pm$13 mV for the PMT-based detector and 
770$\pm$54 mV$\cdot$ns for the SiPM-based array. The error reported is the standard deviation of the projected distribution.
We find that our energy calibration performed as expected for both systems, 
and that the response of the SiPM-based array is significantly more uniform across detector pixels.

\begin{figure}[!htbp]
\centering
	\includegraphics[angle=0,width=\columnwidth]{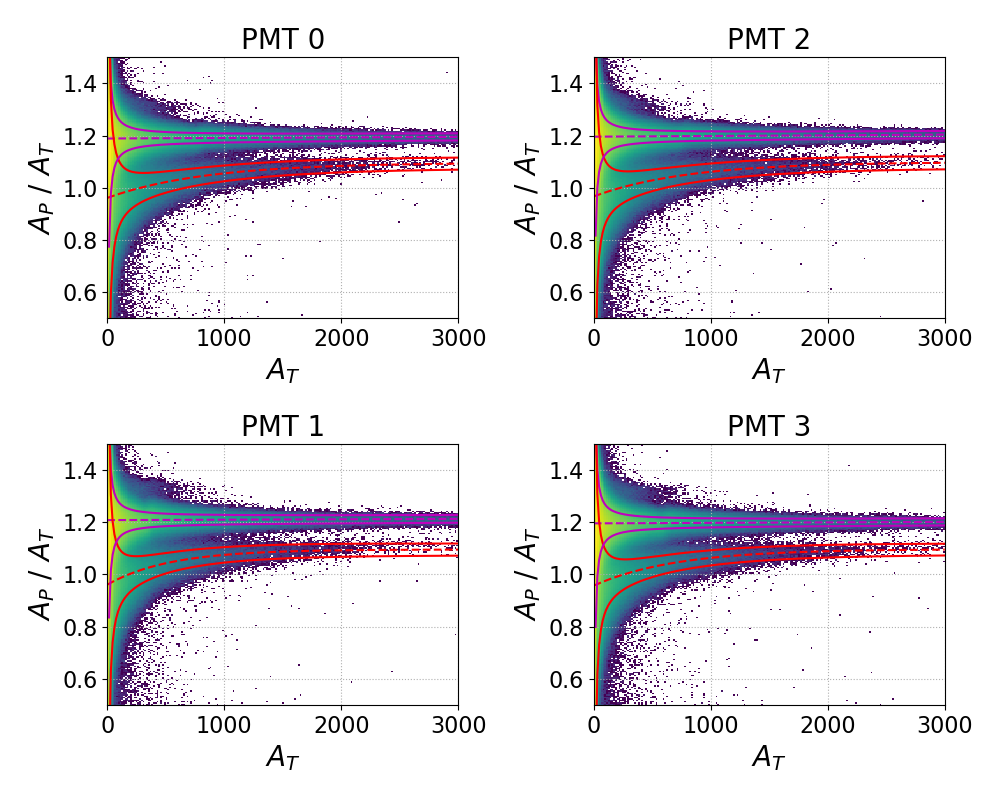}
	\caption{Measured histograms of PSD versus total amplitude value for the four detector PMTs. The dashed and solid lines show the best fit functions (Equations~\ref{eq:blockfitmug} to~\ref{eq:blockfitsig}) for the means and standard deviations of the gamma (magenta) and neutron (red) bands.}
	\label{fig:psdblock} 
\end{figure}

\subsection{Pulse Shape Discrimination}\label{sec:procpsd}

For the PMT-based detector, the PSD values for the four PMTs are given by the prompt amplitude for the given PMT waveform ($A_p$), 
divided by the total amplitude ($A_t$). The PSD measured by each PMT will be used in Section~\ref{sec:eff} to select a neutron population.
The PSD calculated using the sum of the 
four PMT pulses will be used for visualization purposes only. Figure~\ref{fig:psdblock} shows the measured PSD as a function of $A_t$ for each of the 
four PMTs in the PMT-based detector. For each PMT, we bin the $^{252}$Cf dataset into 10-ADC wide increments of $A_t$. 
We combine consecutive increments until at least 5,000 entries are included and then apply double Gaussian fits to the PSD projection in the selceted increments.
This fitting procedure is repeated across the entire range in order to find $A_t$-dependent means and standard deviations 
of the gamma and neutron bands, which are taken from the double Gaussian fits. We then fit simple functional forms to the measured means and standard deviations:  

\begin{align}
\label{eq:blockfitmug} \mu_{\gamma} (A_t) &= a +b A_t \\ 
\label{eq:blockfitmun} \mu_{N}(A_t) &= a+b \exp{cA_t}\\ 
\label{eq:blockfitsig} \sigma_{\gamma,N} ( A_t ) &= a + \frac{b}{A_t} +\frac{c}{A_t^2}.\\
\end{align}
The best fit functions are shown in Figure~\ref{fig:psdblock} and are used to calculate the gamma and neutron probabilities described in Section~\ref{sec:eff}. 

For the SiPM-based array, we characterize the PSD using the ratio of the prompt pulse 
integral to the total pulse integral. The end time of the prompt region is determined by optimizing the figure-of-merit, defined as

\begin{equation}
FoM = \frac{\mu_n-\mu_\gamma}{2.355(\sigma_n+\sigma_\gamma)},
\end{equation}
where $\mu_{n/\gamma}$ is the mean of a Gaussian fit to the neutron and gamma distributions, $\sigma_{n/\gamma}$ is the 
standard deviation of the Gaussian fit, and the factor of 2.355 is for conversion to a FWHM width. The shape of the fission neutron 
energy spectrum favors optimizing the $FoM$ for lower energy depositions in order to maximize overall neutron detection efficiency. 

\begin{figure}[!htbp]
\centering
	\includegraphics[angle=0,width=0.6\columnwidth]{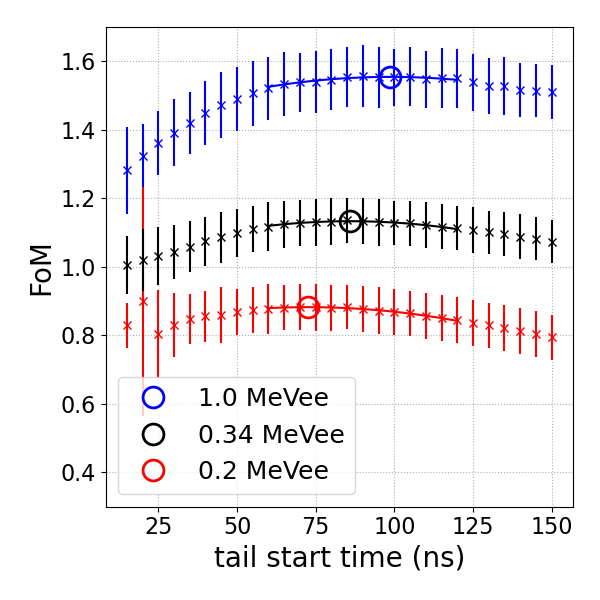}
	\caption{Measurements of the  $FoM$ as a function prompt end time in the SiPM-based array. In this figure, red indicates low-energy measurements 
	(0.2 MeVee), black indicates measurements at 0.34 MeVee, and blue indicates measurements at 1 MeVee. The ``x'' markers show measurements 	
	obtained for the given tail-start time in ns. The error bars indicate the standard deviation of the $FoM$ values for the 64 pixels for each prompt end time. The solid line shows a quadratic fit to measurements near the maximum for each energy selection. Finally, the open circle markers indicate 
	the maximum values of the quadratic fits.  }
	\label{fig:psdcalscan} 
\end{figure}

\begin{figure}[!htbp]
\centering
\begin{subfigure}[h]{0.45\columnwidth} 
			\includegraphics[angle=0,width=\columnwidth]{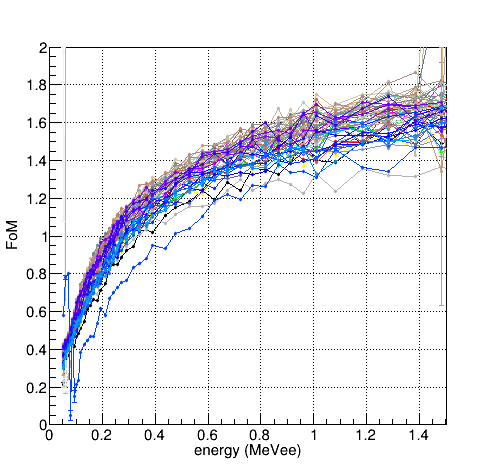}
			\caption{}
		\end{subfigure}	
		\begin{subfigure}[h]{0.45\columnwidth} 
			\includegraphics[angle=0,width=\columnwidth]{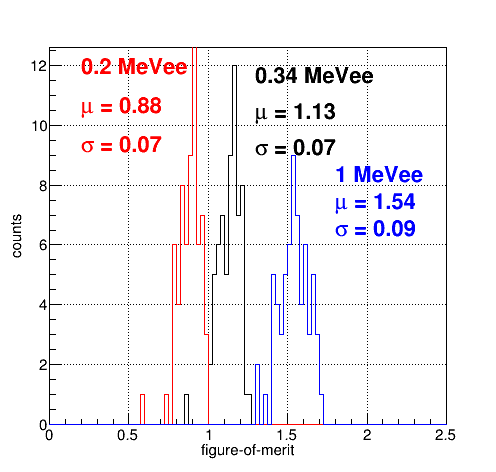}
			\caption{}
		\end{subfigure}

		\caption{(a) Optimized PSD $FoM$ for all 64 pixels in the SiPM-based detector. The high-noise channel mentioned in Section~\ref{sec:exp} is apparent in this panel as the trace with the lowest $FoM$.  (b) The projected $FoM$ for all 64 pixels at 0.2 MeVee (red), 0.34 MeVee (black) and 1 MeVee (blue).}
	\label{fig:figpsd} 
\end{figure}

We scan across prompt end times to identify the value that optimizes the PSD $FoM$ for the SiPM-based array.
For each promt end time selection, we evaluate linear interpolations of the calculated $FoM$ for each pixel in the detector at 0.2~MeVee, 0.34~MeVee, and 1~MeVee.
Figure~\ref{fig:psdcalscan} shows the pixel-wise mean and standard deviation of the evaluated $FoM$ values for each of the prompt end times in this scan. 
We select a prompt end time of 75~ns, as this optimizes the performance at low energy.
The optimized energy-dependent $FoM$ for all pixels in the SiPM-based array is shown in Figure \ref{fig:figpsd}a. 
We report the mean and standard deviation of the $FoM$ across all pixels in the detector in Figure \ref{fig:figpsd}b.

\section{Neutron Detection Efficiency}
\label{sec:eff}

The intrinsic neutron efficiency is defined as the ratio of detected neutrons, $N_d$, to neutrons incident on the detector's face, $N_{i}$.
To calculate $N_{i}$ from the $^{252}$Cf source over all energies, we start with the source activity, $A$, measured at some time $t$
after a survey, to determine the current activity:

\begin{equation}
A = A_0 2^{-t/t_{1/2}},
\end{equation}
where $t_{1/2}$ is the 2.645 year half-life and $A_0$ is the surveyed activity.  
The solid angle coverage ($\Omega$) is calculated by summing the solid angles, $\omega_p$ of each rectangular pixel, $p$, assuming a point source: 

\begin{equation}
\omega_p = \frac{\mathcal{A}\cos{\theta_p}}{ r_p^2} = \frac{a^2h}{ r_p^3},
\end{equation}
 in which, $\mathcal{A}$ is the area of the face of a pixel, including the inter-pixel gap, $a$ is the pixel pitch, $r_p$ is the distance to the center of the face of pixel 
 $p$, $\theta_p$ is the opening angle from the source to the center 
of the pixel, and $h$ is the source distance from the detector face. 
We include the gaps between pixels in this calculation because a neutron that enters the face of the detector through a gap may have a 
trajectory that brings it into the bulk and may interact with the scintillator. The combined solid angle coverage, $\Omega$, of the detector face is given by:

\begin{equation}
\Omega = \sum_{p = 0}^{n-1}\omega_p,
\end{equation}
where $n$ is the total number of pixels. Finally, we account for the 3.09\% spontaneous fission 
branching fraction, $B$, and the average number of neutrons emitted per fission, 
$\overline{\nu}$ = 3.76 \cite{croft}. The total number of incident neutrons expected on the face of the detector for a 
live time $\tau$ (accounting for dead time) is thus

\begin{equation}
N_{i} =  AB\overline{\nu}\tau\frac{\Omega}{4\pi} = A_0 2^{-t/t_{1/2}}\tau B \overline{\nu}\frac{\Omega}{4\pi},
\label{eq:nidef}
\end{equation}

We also estimate the ratio of neutrons detected, $N_d$, per detectable neutron interaction, $N_D$. 
We estimate $N_D$ using Geant4\cite{geant4} simulations of the experimental setup described in Section~\ref{sec:exp}. 
We use Geant4 version 10.7.4 with the QGSP\_BERT\_HP physics list. 
We simulate a number, $S$, of $^{252}$Cf fission neutrons, isotropically emitted from a point source.
We then count the number, $n_D$, of simulated neutrons that undergo at least one elastic scatter with a hydrogen nucleus in the detector volume. 
The ratio of detectable neutrons per emitted $^{252}$Cf neutron is then used to calculate an estimate of $N_D$:
\begin{equation}
N_D= \left( A_0 2^{-t/t_{1/2}}\tau B \overline{\nu} \right)\frac{n_D }{S}.
\end{equation}
The efficiency for detectable neutrons is intended primarily as a relative measure between the two detectors that controls for geometric effects such as trajectory differences and differences in the gap size between pixels.   

To select the detected neutron population, $N_d$, we use a Bayesian likelihood function constructed from the Gaussian fits
to the PSD distribution to calculate gamma and neutron probability metrics, $\hat{P}_n$ and $\hat{P}_{\gamma}$, for each entry in the datatset \cite{brennen}. 
Any deposition with a $\hat{P}_n$ > 99\% at any energy is included in the neutron population, and 
any deposition with a $\hat{P}_{\gamma}$ > 99\% at any energy is included in the gamma population.  
For the PMT-based detector, the probability metric for source $s= \{n, \ \gamma \}$ is constructed from the product of the individual PMT's contribution to the measurement, indexed by $p$:

\begin{equation}
\hat{P}_{s}(E) =  \frac{\displaystyle\prod_{p=0}^4\mathcal{L}_{s,p}(E)}{\displaystyle\prod_{p=0}^4\mathcal{L}_{n,p}(E) + R\displaystyle\prod_{p=0}^4\mathcal{L}_{\gamma,p}(E) + \mathcal{L}_O}.
\label{eq:probpmt}
\end{equation}
where $\mathcal{L}_{s,p}(E)$ are peak-normalized distributions taken from the Gaussian fits to the neutron ($s=n$) and gamma ($s=\gamma$) 
populations and $\mathcal{L}_O = 0.001$ is the likelihood of a third ``other'' classification. The gamma-to-neutron ratio, $R$, is given by the number of gammas divided by the number of neutrons. The gamma and neutron numbers used in this ratio are calculated using a first-pass analysis of the dataset as described above, where we take $R=1$.  

For the SiPM-based detector, the probability metrics are taken from the individually measured distributions for each detector pixel, $p$:
\begin{equation}
\hat{P}_{s,p}(E) =  \frac{\mathcal{L}_{s,p}(E)}{\mathcal{L}_{n,p}(E) + R_p(E)\mathcal{L}_{\gamma,p}(E) + \mathcal{L}_O},
\label{eq:probsipm}
\end{equation}
The energy-dependent gamma-to-neutron ratio for the SiPM-based detector is calculated using the best-fit Gaussian areas:
\begin{equation}
R_p(E) = \frac{A_{\gamma,p}(E)\sigma_{\gamma,p}(E)}{A_{n,p}(E)\sigma_{n,p}(E)}
\end{equation}

\begin{figure}[!htbp]
\centering
		\begin{subfigure}[h]{0.32\columnwidth} 
			\includegraphics[angle=0,width=\columnwidth]{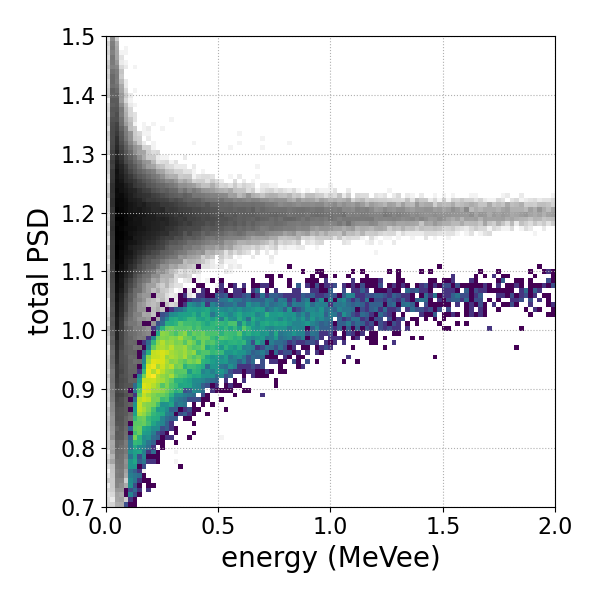}
			\caption{}
		\end{subfigure}
		\begin{subfigure}[h]{0.32\columnwidth} 
			\includegraphics[angle=0,width=\columnwidth]{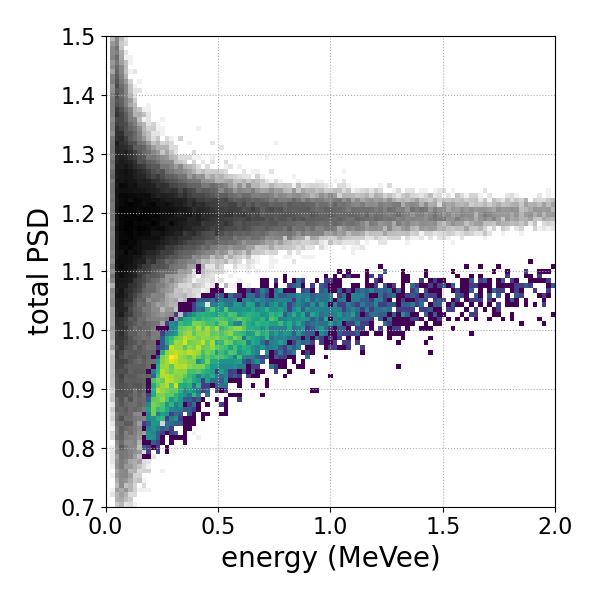}
			\caption{}
		\end{subfigure}
\begin{subfigure}[h]{0.32\columnwidth} 
			\includegraphics[angle=0,width=\columnwidth]{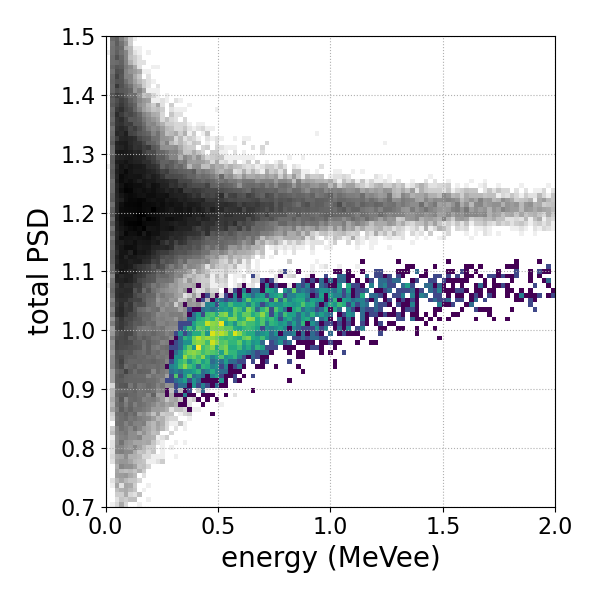}
			\caption{}
		\end{subfigure}	

		\caption{Example neutron selections for pixels in the PMT-based array for a central pixel (a), an edge pixel (b), and a corner pixel (c). The bins in grayscale show events not selected as neutrons, and the blue-green bins indicate selected neutrons. }
	\label{fig:blockneutexamples} 
\end{figure}

\begin{figure}[!htbp]
\centering
\begin{subfigure}[h]{0.32\columnwidth} 
			\includegraphics[angle=0,width=\columnwidth]{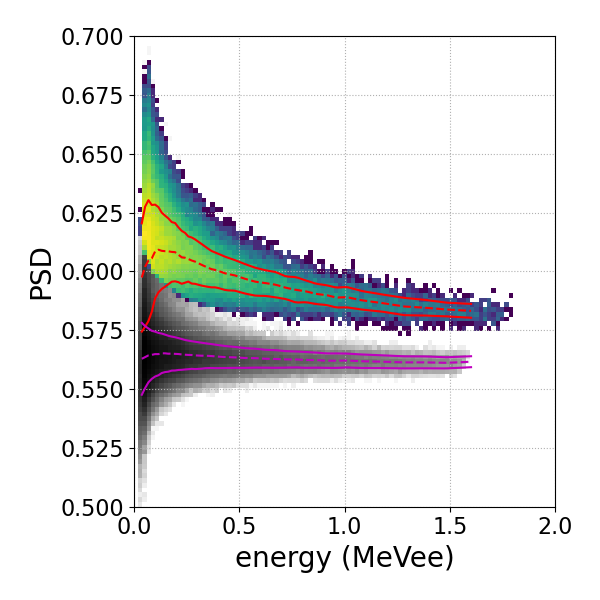}
			\caption{}
		\end{subfigure}
\begin{subfigure}[h]{0.32\columnwidth} 
			\includegraphics[angle=0,width=\columnwidth]{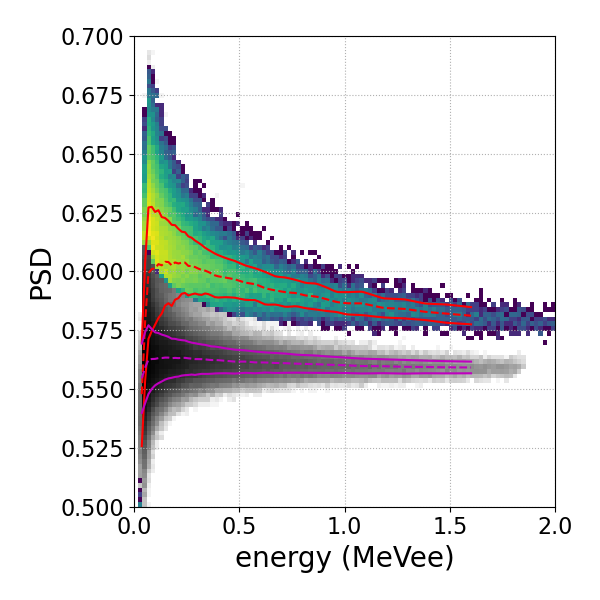}
			\caption{}
		\end{subfigure}
\begin{subfigure}[h]{0.32\columnwidth} 
			\includegraphics[angle=0,width=\columnwidth]{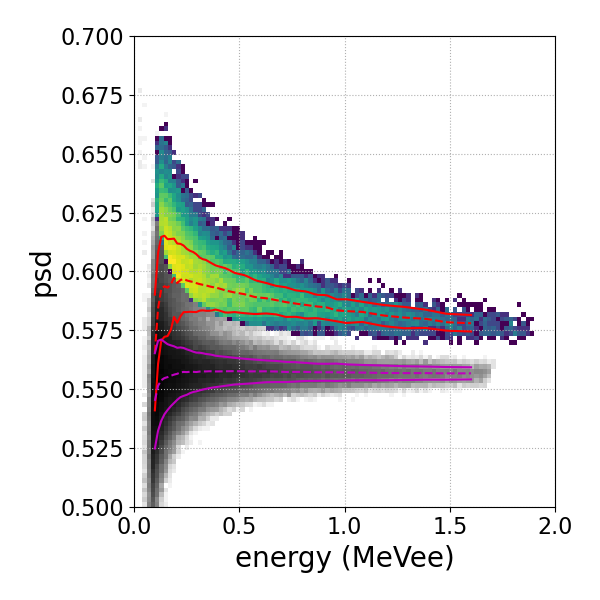}
			\caption{}
		\end{subfigure}	
		
		\caption{Example neutron selections for pixels in the SiPM-based array for a central pixel (left), a corner pixel (center), and a pixel with 
		excess electronic noise (right). The bins in grayscale show events not selected as neutrons, and the blue-green bins indicate selected 
		neutrons. The mean (dashed lines) and 1-sigma widths (solid lines) of the gamma (magenta) and neutron (red) bands for the given pixels 
		are indicated. }
	\label{fig:sipmneutexamples} 
\end{figure}

Figures~\ref{fig:blockneutexamples} and 
\ref{fig:sipmneutexamples} show examples of the neutron selection in various detector pixels in the PMT-based and SiPM-based detectors, 
respectively. The ``total PSD'' measure in Figure~\ref{fig:blockneutexamples} is equal to the sum of the prompt amplitudes across the four 
PMTs, divided by the sum of the total amplitudes. This measure is used for visualization only and is not used in the neutron selection. 
The PSD values in Figure~\ref{fig:sipmneutexamples} are then used for neutron selection in the given pixel. Additionally, the mean and 
1-sigma widths of the gamma and neutron bands for the given pixels are indicated in Figure~\ref{fig:sipmneutexamples}.

In the neutron selection for the PMT-based detector (Figure~\ref{fig:blockneutexamples}), we find an obvious decrease in neutron population 
when comparing a central pixel to an edge pixel, and a central pixel to a corner pixel. The SiPM-based populations show much more 
consistency across pixels. The neutron selection for the worst performing pixel in the SiPM-based detector (pixel number 59) is shown in 
Figure~\ref{fig:sipmneutexamples}c. This channel had significant excess electronic noise and had the highest trigger threshold setting, which corresponds to an effective neutron threshold of 0.135~MeVee. At higher energies, there is no obvious difference between this channel and others. There is a discrepancy in the apparent endpoints of the spectra shown in Figure~\ref{fig:sipmneutexamples}. This is due to the saturation rejection, which excludes pulses greater than 470~mV in aplitude, combined with variations in light collection.

\begin{figure}[!htbp]
\centering
\begin{subfigure}[h]{0.32\columnwidth} 
			\includegraphics[angle=0,width=\columnwidth]{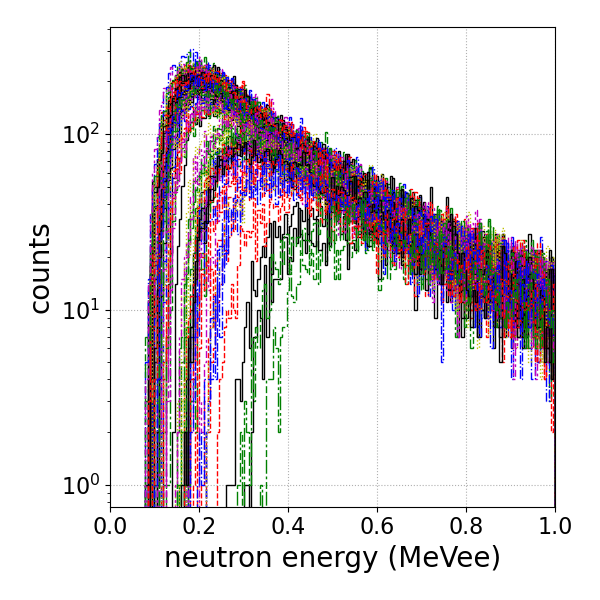}
			\caption{}
		\end{subfigure}	
		\begin{subfigure}[h]{0.32\columnwidth} 
			\includegraphics[angle=0,width=\columnwidth]{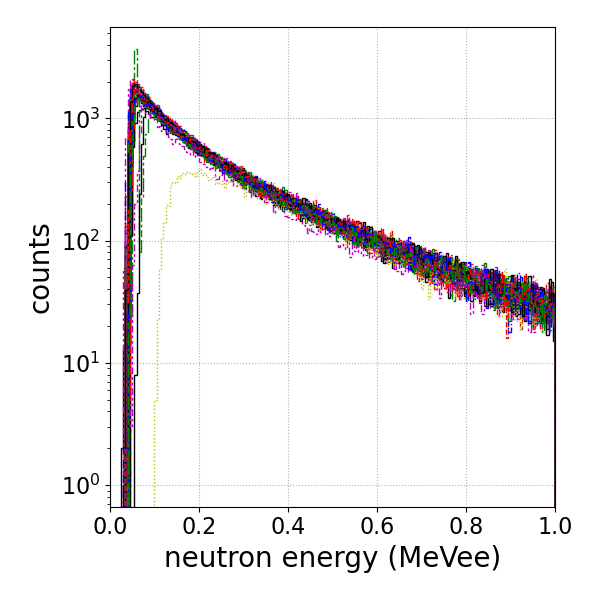}
			\caption{}
		\end{subfigure}

		\caption{Energy spectra for selected neutron pulses for each pixel in the PMT-based detector (left) and SiPM-based detector (right). Each line represents the histogram for a different pixel. }
	\label{fig:neutthresh} 
\end{figure}
 
\begin{figure}[!htbp]
\centering
\begin{subfigure}[h]{0.32\columnwidth} 
			\includegraphics[angle=0,width=\columnwidth]{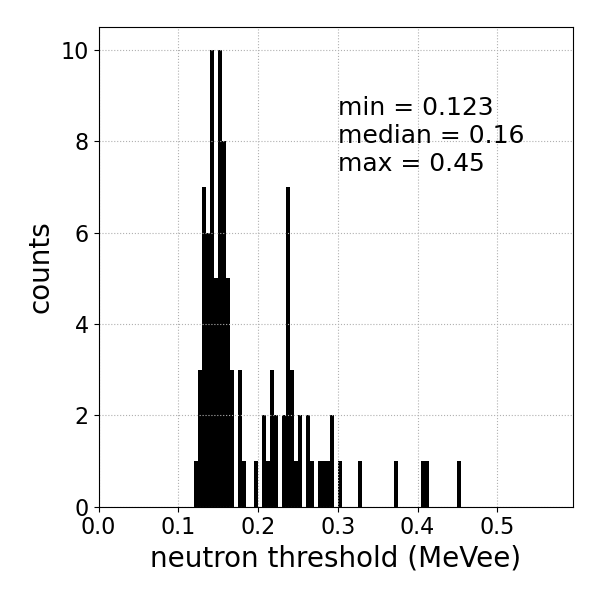}
			\caption{}
		\end{subfigure}	
		\begin{subfigure}[h]{0.32\columnwidth} 
			\includegraphics[angle=0,width=\columnwidth]{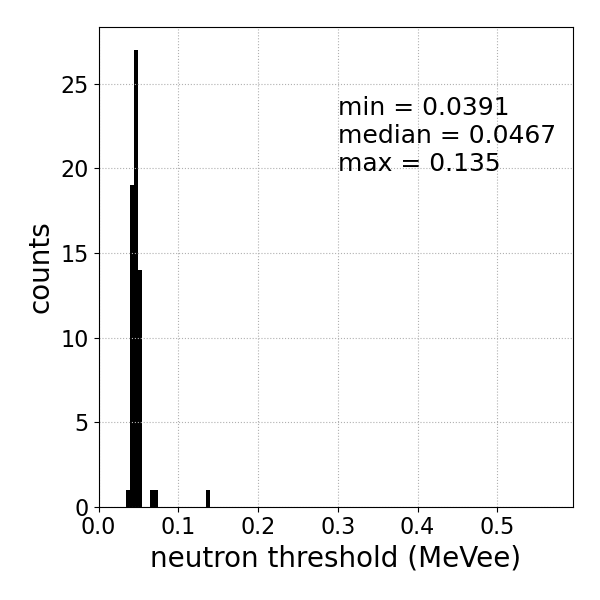}
			\caption{}
		\end{subfigure}

		\caption{Effective energy threshold for neutron events for all pixels in the PMT-based array (left) and the SiPM-based array (right). }
	\label{fig:neutthreshhist} 
\end{figure}

We show the calibrated energy histograms for each pixel in the PMT-based detector and SiPM-based detector in Figure~\ref{fig:neutthresh}. 
The effective energy threshold is measured by fitting the following model function to the spectra shown in Figure~\ref{fig:neutthresh}:

\begin{equation}
f(x) = a\frac{1}{2}\left[ \text{erf} \left( (x-x_0)/b\right)+1\right]\exp\left( -(x-x_0)/c\right)
\end{equation}
where $x_0$, $a$, $b$, and $c$ are fitting parameters. We take $x_0$ to be equal to the effective energy threshold for neutrons. The other fitting parameters  are not used in any further analysis. The histograms of these effective energy thresholds are shown in Figure~\ref{fig:neutthreshhist}. We find that the median threshold for the PMT-based array is 0.16 MeVee, spreading all the way up to 0.45 MeVee, 
while the median threshold for the SiPM-based array is 0.047~MeVee with only a single pixel being higher than 0.1~MeVee. As seen in Figure~\ref{fig:neutthresh}, 
the neutron recoil energy spectrum is highly peaked toward low energy, so the reduced neutron threshold in the SiPM-based array will have a 
large effect on the neutron detection efficiency.

Table~\ref{tab:sourcesetup} contains values for the $^{252}$Cf experimental setup required for calculation of $N_i$ in Equation~\ref{eq:nidef}. 
The live time, $\tau$, is equal to the run time, $t_{run}$, minus the dead time, $t_{dead}$. The dead time for the PMT-based detector is given by
the number of events in the dataset before neutron selection, $N_{total}$, times trigger hold-off period of 2048~ns:
\begin{equation}
t_{dead}\text{(PMT-based)} = N_{total}\cdot (2048 \text{ns}),
\end{equation}
The dead time for the SiPM-based detector varies pixel-to-pixel, so we estimate an average value with a term for dead time in the same pixel due to the trigger hold off time of 1024~ns and a term for dead time in all other pixels due to the coincidence rejection:
\begin{equation}
t_{dead}\text{(SiPM-based)} = \frac{1}{64}\cdot N_{total}(1024 \text{ns}) + \frac{63}{64}\cdot N_{total}(50 \text{ns}).
\end{equation}

\begin{center}
\begin{tabular}{||c || c || c || c || c || c ||} 
 \hline
 Detector & $h \ \pm \ \sigma$ (cm) & $\Omega$ (sr) & $A$ ($\mu$Ci)  & $t_{run}$ (s) & $t_{dead}$ (s) \\ [0.5ex] 
 \hline\hline
 PMT-based & 142 $\pm$ 1 & 0.005764 & 154.8  & 21600.0 & 37.5 \\
 \hline
 SiPM-based & 40 $\pm$  1 & 0.01603 & 155.0  & 14400.0 & 1.7 \\
 \hline 
\end{tabular}
\captionof{table}{Table of values for the $^{252}Cf$ experimental setup required for calculation of $N_i$ in Equation~\ref{eq:nidef}. 
The live time, $\tau$, is equal to the run time, $t_{run}$, minus the dead time, $t_{dead}$.}
\label{tab:sourcesetup}
\end{center}

Table~\ref{tab:resultsintrin} contains the results for the intrinsic neutron detection efficiency measurement for both the PMT-based detector and 
the SiPM-based detector. The value of $N_d$ is the number of neutron events for which the calculated neutron probability is >99\%. The 
statistical uncertainty is taken to be $\sqrt{N_d}$. For the PMT-based detector, we estimate the systematic error on $N_d$ using the fractional deviation 
about the best-fit line in Figure~\ref{fig:blockpromptopt}. For the SiPM-based detector, we estimate the systematic error on $N_d$ using:

\begin{equation}
\sigma_{syst} = (N_{p>0.984} - N_{p>0.994})/2,
\end{equation}
where $N_{p>0.984}$ is the number of detected pulses with neutron probability >98.4\%, and $N_{p>0.994}$ is the number of detected pulses 
with neutron probability >99.4\%. The values of 98.4\% and 99.4\% are somewhat arbitrary, but are motivated by the uncertainty in the low-energy 
PSD $FoM$ of the SiPM-based detector shown in Figure~\ref{fig:figpsd}. The cumulative distribution function (CDF) for a standard normal random 
variable is equal to  99\% at a value of $X_{0.99} = 2.327$. This is roughly analogous to the distance in normalized PSD space a pulse needs to be 
from the gamma mean in order to be considered a neutron. The PSD $FoM$ is a measure of the distance in PSD space between the neutron and 
gamma means, so we use the fractional uncertainty in the PSD $FoM$ as a measure of how well distances in PSD space are known. We 
decrease $X_{0.99}$ by the fractional uncertainty in $FoM$ and use the adjusted CDF value (98.4\%) for the lower bound for our $\sigma_{syst}$ 
calculation. Similarly, we then increase $X_{0.99}$ by the fractional uncertainty in PSD $FoM$ and use the adjusted CDF value (99.4\%) for the upper
bound. We quote the uncertainty on $N_i$ due to the source placement error described in 
Section~\ref{sec:exp}. The intrinsic neutron detection efficiency ($N_d/N_i$) is reported in Table \ref{tab:resultsintrin}, along with the error propagated 
from the stated sources.

\begin{center}
\begin{tabular}{||c || c | c | c | c ||} 
 \hline
 Detector & $N_d$ $\pm$  $\sigma_{count}$ $\pm$  $\sigma_{syst}$ & $N_i$ $\pm$  $\sigma_{syst}$  &   $N_d/N_i$ $\pm$  $\sigma$ \\ [0.5ex] 
 \hline\hline
 PMT-based & 1112031 $\pm$ 1055 $\pm$ 4645 & 6594197 $\pm$ 26773 & 0.169 $\pm$ 0.001 \\ 
 \hline
 SiPM-based & 3678888 $\pm$ 1918 $\pm$ 100618 & 12178050 $\pm$ 175103 & 0.302 $\pm$ 0.009 \\ 
 \hline 
\end{tabular}
\captionof{table}{Table of results for the intrinsic neutron detection efficiency ($N_d/N_i$).}
\label{tab:resultsintrin}
\end{center}

Table~\ref{tab:resultsintera} contains the results for the neutron detection efficiency per unique neutron interaction for both the PMT-based detector and the SiPM-based detector. In contrast to the 
calculated $N_i$ value, the measured value of $N_D$ has uncertainty due to counting. The fractional systematic uncertainty for $N_D$ is calculated in the same way as that of $N_i$.

\begin{center}
\begin{tabular}{||c || c | c | c ||} 
 \hline
 Detector & $N_d$ $\pm$ $\sigma_{count}$ $\pm$  $\sigma_{syst}$ & $N_D$ $\pm$  $\sigma_{count}$ $\pm$  $\sigma_{syst}$ &  $N_d/N_D$ $\pm$  $\sigma$ \\ [0.5ex] 
 \hline\hline
 PMT-based & 1112031 $\pm$ 1055 $\pm$ 4645 & 3545600 $\pm$ 7132 $\pm$ 14395 & 0.314 $\pm$ 0.002 \\ 
 \hline
  SiPM-based & 3678888 $\pm$ 1918 $\pm$ 100618 & 5628554 $\pm$ 7346 $\pm$ 80931 & 0.654 $\pm$ 0.020 \\ 
 \hline 
\end{tabular}

\captionof{table}{Table of results for the neutron detection efficiency per detectable ($N_d/N_D$).}
\label{tab:resultsintera}
\end{center}

\begin{figure}[!htbp]
\centering
\begin{subfigure}[h]{0.4\columnwidth} 
			\includegraphics[angle=0,width=\columnwidth]{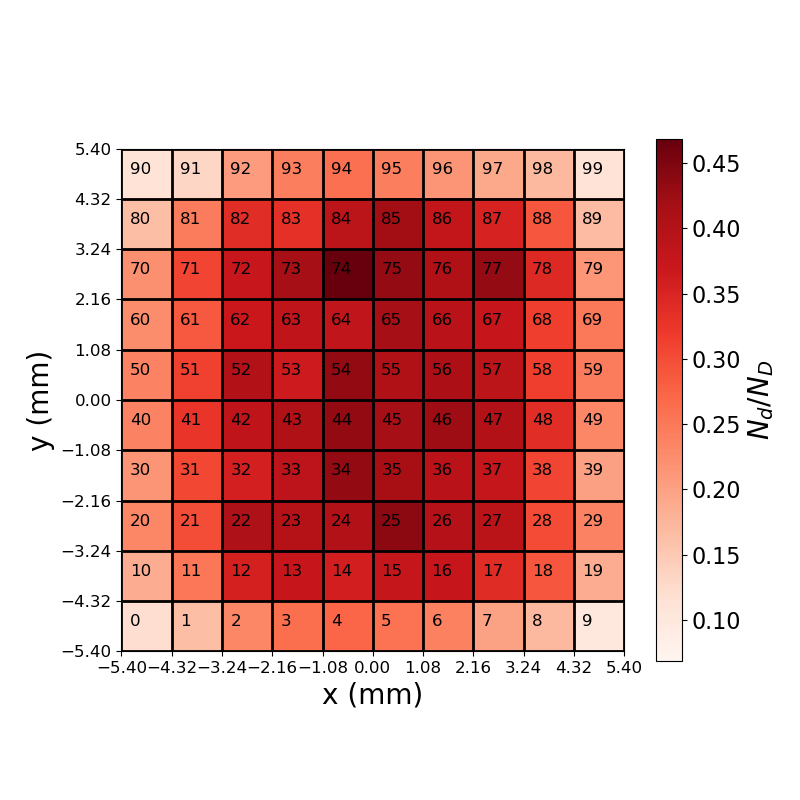}
			\caption{}
		\end{subfigure}	
		\begin{subfigure}[h]{0.4\columnwidth} 
			\includegraphics[angle=0,width=\columnwidth]{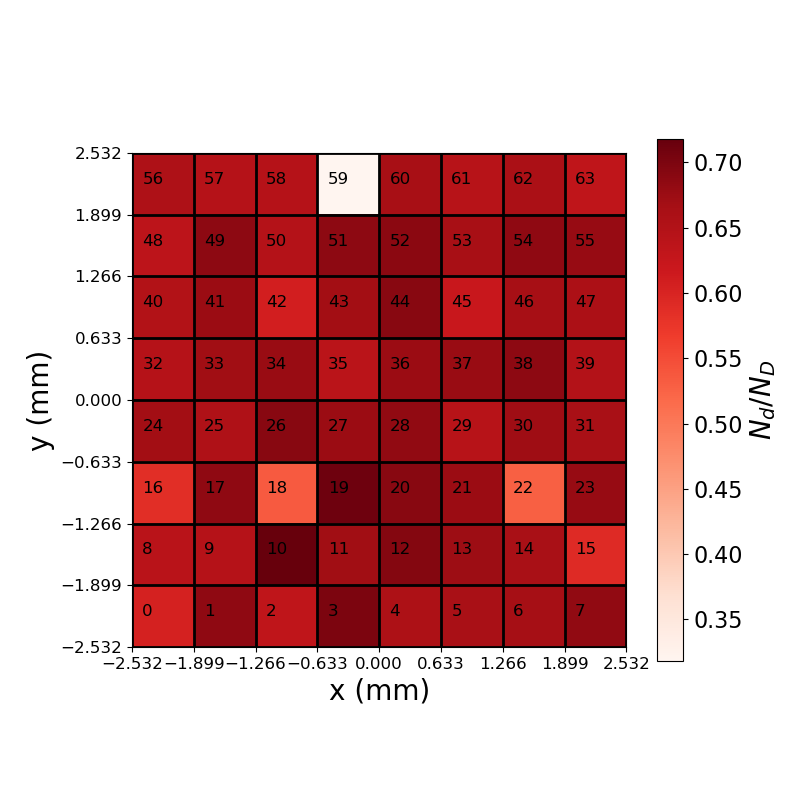}
			\caption{}
		\end{subfigure}

		\caption{Efficiency maps for the PMT-based detector (left) and for the SiPM-based detector (right). Each grid-square indicates a separate pixel, the index of which is noted in the square. 
		The color scale indicates the value of $N_d/N_D$ for the given pixel. The width of the color scale range for each plot is equal to 0.4, with the upper limit equal to the maximum pixel value for the given detector. }
	\label{fig:effmaps} 
\end{figure}

\begin{figure}[!htbp]
\centering
\begin{subfigure}[h]{0.4\columnwidth} 
			\includegraphics[angle=0,width=\columnwidth]{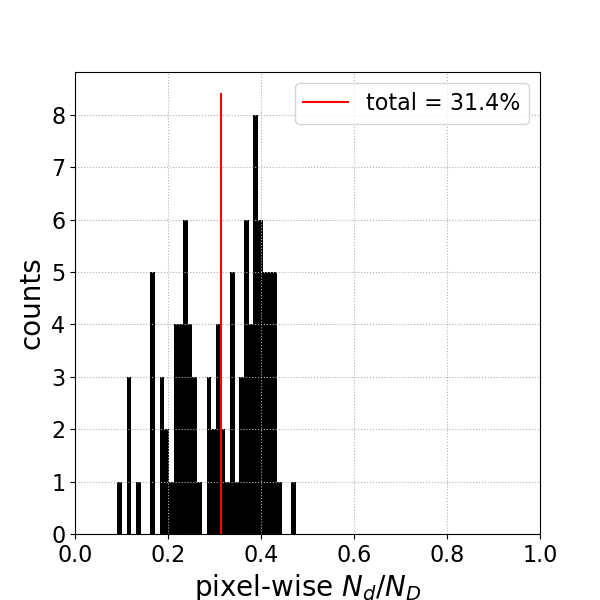}
			\caption{}
		\end{subfigure}	
		\begin{subfigure}[h]{0.4\columnwidth} 
			\includegraphics[angle=0,width=\columnwidth]{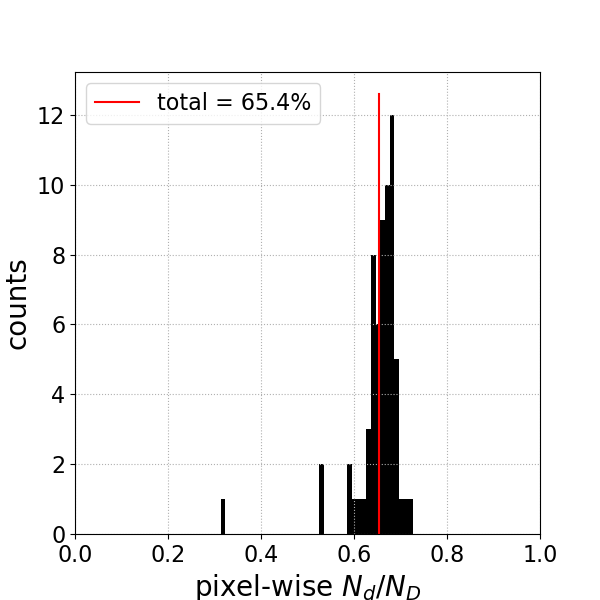}
			\caption{}
		\end{subfigure}

		\caption{Histograms of the pixel-wise values of $N_d/N_D$ for the PMT-based detector (left) and the SiPM-based detector (right). The vertical red line indicates the efficiency of
		the entire detector, listed in Table~\ref{tab:resultsintera}.}
	\label{fig:pixinteffhist} 
\end{figure}

In Figures~\ref{fig:effmaps}  and~\ref{fig:pixinteffhist}, we break $N_d/N_D$ up into pixel-wise values. We find, as expected, that the efficiency is much more uniform in the the SiPM-based array than it is in the PMT-based array.  

\section{Conclusion}

We have examined the intrinsic neutron detection efficiency for both a PMT-based scintillator array using Anger logic readout and a SiPM-based array 
with 1-to-1 scintillator/photodetector coupling. The use of SiPMs with 1-to-1 coupling greatly improves both the pixel uniformity and the absolute performance of the detector. 
We find that the intrinsic neutron detection efficiency is (16.4 $\pm$ 0.2)\% for the PMT-based detector, versus (30.2 $\pm$ 1.7)\% for the SiPM based detector. 
We also report estimated values of neutron detection efficiency per neutron interaction using a Geant4 simulation of the two detectors. We find that the 
neutron detection efficiency per detectable neutron is (31.4 $\pm$ 0.5)\% for the PMT-based array, versus (65.4 $\pm$ 3.7)\% for the SiPM-based array. This 
means that for the same mass of detector material, a SiPM-based pixelated readout would detect 2.1 times the number of neutrons as a PMT-based detector that uses Anger logic. 

If the existing detectors were located at the same distance from a neutron source, the SiPM-based detector would be expected to detect about 40\% of the neutrons that the PMT-based 
detector despite only having about 20\% of the active volume. A 2x2 set of SiPM-based detectors would have nearly identical coverage as a single PMT-based detector, and might be used 
as a drop-in replacement. The resulting combined detector array would have 60\% finer pixelation than the PMT-based detector and about 60\% higher absolute efficiency.

\section{Acknowledgments}

The authors would like to thank Steven Hammon from Sandia National Laboratories, who performed the layout of the 64 channel SOUT readout board and 
Kyle Weinfurther, also of Sandia National Laboratories, for performing an independent review of this publication.
We would also like to thank Matthew Blackston, Paul Hausladen, and Jason Newby for their prior work in design and algorithm development for 
the PMT-based detector. 
Paul Hausladen also worked with with Agile Technologies to provide the SiPM-based scintillator array. 
Paul Hausladen and Jason Newby are with Oak Ridge National Laboratory, and Matthew Blackston was formerly with Oak Ridge National Laboratory.

Finally, we thank the US DOE National Nuclear Security Administration, 
Office of Defense Nuclear Nonproliferation Research and Development for funding this work. 
Sandia National Laboratories is a multimission laboratory managed and operated by National Technology and Engineering Solutions of Sandia, LLC, a wholly owned subsidiary of Honeywell 
International, Inc., for the U.S. Department of Energy's National Nuclear Security Administration under contract DE-NA0003525. This paper describes objective technical results and analysis. 
Any subjective views or opinions that might be expressed in the paper do not necessarily represent the views of the U.S. Department of Energy or the United States Government. 
Document Release Number SAND2023-09226O.

\bibliographystyle{../../../LatexTools/IEEE_style/IEEEtran}

\begin{thebibliography}{1}
\bibitem{anger} H. O. Anger. {\it Review of Scientific Instruments } {\bf 29 } (1958) 27-33 
\bibitem{newby2} R. J. Newby, P.A. Hausladen, M. A. Blackston, J. F. Liang. {\it ORNL/TM-2013/82: https://doi.org/10.2172/1128961}
\bibitem{zappala} G. Zappalà {\it et al. Journal of Instrumentation } {\bf 11} (2016) P08014
\bibitem{nakamura} K. Nakamura {\it et al.  Nucl. Inst. and Meth. in Phys. Res. A.} {\bf 623} (2010) 276-278
\bibitem{romeo} G. Romeo {\it et al. Nucl. Inst. and Meth. in Phys. Res. A.} {\bf 826} (2016) 31-38
\bibitem{zhao1} Zhixiang Zhao {\it et al.} {\it IEEE Trans. Nucl. Sci.} {\bf 64 (2)}  (2017) 820 
\bibitem{zhao2} Zhixiang Zhao {\it et al.}   {\it IEEE Trans. Nucl. Sci.} {\bf 68 (9)} (2019) 3200
\bibitem{zheng} Jiajun Zheng {\it et al. IEEE Trans. Nucl. Sci.} {\bf 64 (6)} (2017) 1401 
\bibitem{newby1} J. Newby, P. Hausladen, and M.A Blackston. "Position-Sensitive Organic Scintillators for Nuclear Material Accountancy" { \it Symposium on International Safeguards: Linking Strategy, Implementation and People} (2014)  https://www.osti.gov/servlets/purl/1163163
\bibitem{giha} Nathan P. Giha {\it IEEE Nucl. Sci. Symp. Atlanta, Ga. } DOI: 10.1109/NSSMIC.2017.8532622 (2015)
\bibitem{dt5730} "DT5730/DT5725 User Manual, Rev. 8"  {\it CAEN S.p.A.} (2023)
\bibitem{compass} "User Manual UM5960 CoMPASS Multiparametric DAQ Software for Physics Applications, Rev. 20" {\it CAEN S.p.A.} (2020)
\bibitem{v1725} "V1730/VX1730 \& V1725/VX1725 User Manual, Rev. 7"  {\it CAEN S.p.A.} (2023)
\bibitem{root}  R. Brun and F. Rademakers.  {\it Nucl. Inst. and Meth. in Phys. Res. A} {\bf 389 } (1997) 81-86
\bibitem{numpy} C. R. Harris {\it et al.} {\it Nature} {\bf 585 (7825)}{\bf } (2020) 357-362
\bibitem{scipy} P. Virtanen {\it et al.} {\it Nature Methods} {\bf 17} (2020) 261-272
\bibitem{mpl} J. D. Hunter {\it Computing in Science \& Engineering} {\bf 9 (3)} (2007) 90-95
\bibitem{uproot} J. Pivarski {\it et al.} "Uproot v5.0.11" {\it Zenodo} (2023) {\it https://doi.org/10.5281/zenodo.8239801}  
\bibitem{klein} H. Klein and S. Neumann. {\it Nucl. Inst. and Meth. in Phys. Res. A.} {\bf 476} (2002) 132-142
\bibitem{sweany} M. Sweany {\it et al. Nucl. Inst. and Meth. in Phys. Res. A } {\bf 927 }(2019) 451-462
\bibitem{croft} S. Croft, A. Favalli, and R. D. McElroy Jr. { \it Nucl. Inst. and Meth. in Phys. Res. A}{\bf 954} (2020) 161605
\bibitem{geant4} J. Allison {\it et al.} ``Geant4 developments and applications" {\it IEEE Transactions on Nuclear Science} {\bf 53} (2006) 270-278
\bibitem{brennen} J. Brennan, E. Brubaker, M. Gerling, P. Marleau, M. Monterial, A. Nowack, P. Schuster, B. Sturm, M. Sweany. { \it Nucl. Inst. and Meth. in Phys. Res. A}{\bf 877} (2018) 375-383


\end{thebibliography}

\end{document}